\documentclass[12pt]{article}
\usepackage{amsfonts}

\usepackage{amsmath}
\usepackage{amsthm}
\usepackage{amssymb}

\def\be{\begin{equation}}
\def\ee{\end{equation}}
\def\bea{\begin{eqnarray}}
\def\eea{\end{eqnarray}}
\def\al{\alpha}
\def\ga{\gamma}

\def\1{\'{\i}}           
                
\newcommand{\bq}{\mathbf{q}}
\newcommand{\bp}{\mathbf{p}}
\def\rrr{{\cal R}}

\def\dd{{\rm d}}

\def\back{\!\!\!\!\!\!\!\!\!\!}
\def\jp{J_+}
\def\jm{J_-}
\def\jj{J_3}
\def\otra{b}

 \def\kk{K}

\def\la{\lambda}
 
\def\d{{\rm d}}

\def\g{g}

\def\R{{\mathbb R}} 
 
\def\k{\kappa}
\def\>#1{{\mathbf#1}}

\def\otra{b}
\def\co{\Delta}  
\def\pois#1#2{\left\{ {#1},{#2} \right\}}

\def\la{\lambda}
\def\pot{{\cal U}}

\def\tfrac#1#2{ {\scriptstyle { \frac {#1}{#2}}}}         
\def\pois#1#2{\left\{ {#1},{#2} \right\}}

\def\funciona{{\cal G}}
\def\funcionb{{\cal F}}

\def\sss{{\cal S}}

\newcommand{\bla}{\boldsymbol{\lambda}}

\begin{document}

\thispagestyle{empty}

\hfill \

\ 
\vspace{0.5cm}

\noindent
 {\Large{\bf
 {(Super)integrability from coalgebra symmetry: \vspace{0.2cm}
 \\ formalism and applications
  }}}

\bigskip

\medskip

\begin{center}

 {A. Ballesteros$^1$, A. Blasco$^1$,  F.J. Herranz$^1$, F. Musso$^1$ and
O. Ragnisco$^{2}$}

\end{center}

\noindent
 {$^1$ Departamento de F\1sica, Universidad de Burgos, 
E-09001 Burgos, Spain}

\noindent
 {$^2$ Dipartimento di Fisica,   Universit\`a di Roma Tre and Instituto
Nazionale di Fisica Nucleare sezione di Roma Tre,  
Via Vasca Navale 84, 
I-00146 Roma, Italy}
 
  \medskip

\noindent
e-mails: {angelb@ubu.es, ablasco@ubu.es, fjherranz@ubu.es, fmusso@ubu.es,\\ ragnisco@fis.uniroma3.it}

  \medskip

\begin{abstract} The coalgebra approach to the construction of classical integrable systems from Poisson coalgebras  is reviewed, and the essential role played  by symplectic realizations in this framework is emphasized.
Many examples of Hamiltonians with either undeformed or $q$-deformed coalgebra symmetry are given, and their Liouville superintegrability is discussed. Among them, (quasi-maximally) superintegrable systems on $N$-dimensional curved spaces of nonconstant curvature are analysed in detail. Further generalizations of the coalgebra approach that make use of comodule and loop algebras are presented. The generalization of such a coalgebra symmetry framework to quantum mechanical systems is straightforward.
\end{abstract}


\section{Introduction}

Poisson coalgebras are Poisson algebras
endowed with a compatible coproduct structure. The aim of this contribution is to provide a self-contained review of a recently introduced symmetry approach in which Poisson coalgebras play an essential role as `hidden' dynamical symmetries underlying
the (super)integrability properties of a wide class of $N$-dimensional ($N$D)
classical Hamiltonian systems.  Within this construction,
once a  symplectic realization of the coalgebra is given, their generators 
play the role of  dynamical symmetries of the Hamiltonian --which is written as a function of them-- while   the coproduct map (coalgebra structure)  is used to `propagate' the integrability to any arbitrary
dimension. 

From this so-called coalgebra approach,  many well-known 
(super)integrable systems have been recovered,  and some integrable
deformations for them  as well as new $N$D integrable systems    have also
 been    obtained (see
\cite{BCR}--\cite{alfonso} and references therein).
As a remarkable application, this framework   has  been recently
used to introduce integrable Hamiltonians describing geodesic flows on
spaces with  either constant or nonconstant curvature, and
(super)integrable potential terms preserving  the coalgebra symmetry can also be considered on such
spaces~\cite{plb}--\cite{2photon}. 

Although in this contribution we shall concentrate on classical mechanical systems obtained from
(commutative) Poisson coalgebras, we stress that all the (super)integrability properties of the coalgebra
symmetric systems that we are going to describe are preserved at the quantum mechanical level by using the corresponding
(noncommutative) operator coalgebras. In this way, the coalgebra approach has also been  used, for
instance, to solve in \cite{MRjmp,MRjpa}  the quantum Calogero--Gaudin system \cite{C&vD}, as well as
supersymmetric generalizations of this model and of its $q$-deformations, which have   been  constructed
by starting from the underlying supersymmetric coalgebra structures \cite{MRBrei, ospGaudin}.

The paper is organized as follows. In section 2 we present in a complete and self-contained way the general coalgebra approach to integrable Hamiltonians, that will be illustrated in section 3 by applying it to three relevant Poisson coalgebras, namely  $\frak{sl}(2,\R)$, its non-standard quantum deformation,   $\frak{sl}_z(2,\R)$ and the two-photon Poisson-coalgebra $h_6$. Many physically interesting Hamiltonian systems will be explicitly constructed on these and other coalgebras in sections 4, 5, 6 and 7. Finally, some possible generalizations of the coalgebra formalism will be sketched in section 8.


\section{Hamiltonian systems on Poisson coalgebras}

First of all, let us fix the terminology concerning (Liouville) integrability and superintegrability (see, for instance,~\cite{AKN97}).  An
 $N$D  Hamiltonian
$H^{(N)}$ is called {\em completely integrable} if there exists
a set of
$(N-1)$ globally defined and functionally independent constants of the motion in involution that
Poisson-commute with $H^{(N)}$. The  Hamiltonian will be called     {\it maximally superintegrable} (MS) if there exists
a set of
$(2N-2)$ globally defined  functionally independent constants of the motion  
Poisson-commuting with $H^{(N)}$;  among them, at least two
different subsets of
$(N-1)$ constants in involution can be found.
Finally, 
a Hamiltonian system will be  called  {\it
quasi-maximally superintegrable} (QMS) if it has  
$(2N-3)$ independent integrals with the abovementioned properties.

We also recall that a   coalgebra  $(A,\Delta)$ is a (unital, associative) algebra
$A$ endowed with a  coproduct map~\cite{Dri,CP}:
$$
\Delta: A
\rightarrow A\otimes A
$$ 
which is  coassociative 
$$
(\Delta \otimes {\rm id}) \circ \Delta=({\rm id} \otimes \Delta) \circ \Delta 
$$
that is, the following diagram is commutative:
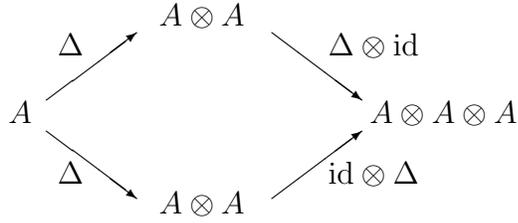
\begin{center}
\setlength{\unitlength}{1cm}
\begin{picture}(6,4)
\put(5,2){$A \otimes A \otimes A$}
\put(.2,2){$A$}
\put(2.2,.8){$A \otimes A$}
\put(2.2,3.3){$A \otimes A$}
\put(3.7,3.2){\vector(4,-3){1.2}}
\put(3.7,1){\vector(4,3){1.2}}
\put(.7,1.9){\vector(4,-3){1.2}}
\put(.7,2.3){\vector(4,3){1.2}}
\put(0.85,2.9){$\Delta$}
\put(0.85,1.2){$\Delta$}
\put(4.45,2.9){$\Delta \otimes  {\rm id}$}
\put(4.45,1.2){$ {\rm id} \otimes \Delta$}
\end{picture}
\end{center}

Due to the coassociativity property, the  comultiplication $\Delta$ provides a `two-fold way' for the
definition of the objects on
$A\otimes A\otimes A$ that,  as we shall explain in section 2.2, will be deeply connected with {\em superintegrability} properties.

Let us now summarize the general construction of~\cite{BR}. Let $(A,\Delta)$ 
be a Poisson coalgebra with $l$ {generators}
$X_i$
$(i=1,\dots,l)$, 
and $r$ functionally independent {Casimir} functions ${\cal{C}}_j(X_1,\dots,X_l)$ 
(with $j=1,\dots,r$). The {coassociative coproduct} $\Delta\equiv\Delta^{(2)}$
is  a {Poisson map} with respect to the usual
Poisson bracket on $A\otimes A$:
$$
\pois{X_i\otimes X_j}{X_r\otimes X_s}_{A\otimes  A}=
\{X_i, X_r\}_A\otimes X_j  X_s +
 X_i  X_r \otimes \{X_j, X_s \}_A  .
$$
Then, the {$m$-th coproduct} map $\co^{(m)}(X_i)$ 
\be
\co^{(m)}:A\rightarrow A\otimes A\otimes \dots^{m)}\otimes A 
\ee
can be defined  by applying recursively the
coproduct $\co^{(2)}$ in the form
\be
\co^{(m)}:=({\rm id}\otimes {\rm id}\otimes\dots^{m-2)}\otimes {\rm id}\otimes
\co^{(2)})\circ\co^{(m-1)}.
\label{fl}
\ee
Such an induction ensures that $\co^{(m)}$ is also a Poisson map.

\begin{table}[t]
{\footnotesize
 \noindent
\caption{{Functions obtained by applying the coproduct
map.}}
\label{table1}
\medskip
\noindent\hfill
$$
\begin{tabular}{cccccccc}
\hline
\phantom{{\Large $\!\!\!\!i_{I_I}$}}{\it $X_1$}&
{\it $X_2$}&
{\it $\dots$}&
{\it $X_l$}&
{\it ${\cal{C}}_1$}&
{\it ${\cal{C}}_2$}&
{\it $\dots$}&
{\it ${\cal{C}}_r$}\\
\hline
\phantom{{\huge i}}$\Delta^{(2)}(X_1)$&$\Delta^{(2)}(X_2)$&$\dots$&
$\Delta^{(2)}(X_l)$&$\Delta^{(2)}({\cal{C}}_1)$&$\Delta^{(2)}({\cal{C}}_2)$&$\dots$&$\Delta^{(2)}({\cal{C}}_r)$\\
\phantom{{\huge i}}$\Delta^{(3)}(X_1)$&$\Delta^{(3)}(X_2)$&$\dots$&
$\Delta^{(3)}(X_l)$&$\Delta^{(3)}({\cal{C}}_1)$&$\Delta^{(3)}({\cal{C}}_2)$&$\dots$&$\Delta^{(3)}({\cal{C}}_r)$\\
$\vdots$&$\vdots$&\ &
$\vdots$&$\vdots$&$\vdots$& &$\vdots$
\\
\phantom{{\Large $\!\!\!\!i_{I_I}$}}$\Delta^{(m)}(X_1)$&
$\Delta^{(m)}(X_2)$&$\dots$&
$\Delta^{(m)}(X_l)$&$\Delta^{(m)}({\cal{C}}_1)$&$\Delta^{(m)}({\cal{C}}_2)$&$\dots$&$\Delta^{(m)}({\cal{C}}_r)$\\
$\vdots$&$\vdots$&\ &
$\vdots$&$\vdots$&$\vdots$& &$\vdots$\\
\phantom{{\Large $\!\!\!\!i_{I_I}$}}$\Delta^{(N)}(X_1)$&
$\Delta^{(N)}(X_2)$&$\dots$&
$\Delta^{(N)}(X_l)$&$\Delta^{(N)}({\cal{C}}_1)$&$\Delta^{(N)}({\cal{C}}_2)$&$\dots$&$\Delta^{(N)}({\cal{C}}_r)$\\
\hline
\end{tabular}
$$
\hfill}
\end{table}

In this way, we can construct the set of functions shown in table \ref{table1}. 
From them, given a smooth function ${\cal{H}}(X_1,\dots,X_l)$,
the $N$-sites Hamiltonian is defined as the $N$-th coproduct of 
${\cal{H}}$:
\be
H^{(N)}:=\co^{(N)}({\cal{H}}(X_1,\dots,X_l))=
{\cal{H}}(\co^{(N)}(X_1),\dots,\co^{(N)}(X_l)).
\label{htotg}
\ee
From~\cite{BR} it can be proven that the set of $r\cdot N$ functions
$(m=1,\dots,N; j=1,\dots,r)$
\be
C^{(m)}_j:= \co^{(m)}({\cal{C}}_j(X_1,\dots,X_l))=
{\cal{C}}_j(\co^{(m)}(X_1),\dots,\co^{(m)}(X_l)) ,
\label{Ctotg}
\ee
such that $C^{(1)}_j={\cal{C}}_j$, 
Poisson-commute with the Hamiltonian
\be
\pois{C^{(m)}_j}{H^{(N)}}_{A\otimes
A\otimes\dots^{N)}\otimes A}=0 
\label{za1}
\ee
and all these constants are, by construction, in involution:
\be
\pois{C^{(m)}_i}{C^{(n)}_j}_{A\otimes
A\otimes\dots^{N)}\otimes A}=0 \qquad  m,n=1,\dots,N\quad
i,j=1,\dots,r.
\label{cor3}
\ee

This construction can be applied to {\em any} Poisson coalgebra. But there are two relevant families of Poisson coalgebras that will constitute the core of the integrable systems presented in this paper:

\begin{itemize} 

\item {\em Lie--Poisson algebras}. The Poisson analogue of any Lie algebra   with generators $X_i$
$(i=1,\dots,l)$
is a coalgebra when endowed with the (primitive) coproduct map~\cite{Tjin}
\begin{equation}
\co(X_i)= X_i\otimes 1+1\otimes X_i \qquad \Delta(1)=1\otimes 1. 
\label{standardcop}
\end{equation}
Equivalently, this means that Lie algebras are cocommutative Hopf algebras.

\item {\em  Poisson analogues of quantum
algebras and groups}~\cite{Dri,CP, Tjin}. These are also (deformed) coalgebras
$(A_z,\Delta_z)$, where $z$ is the quantum deformation parameter $(q={\rm e}^z)$, and the deformed coproduct gives rise to a noncocommutative Hopf algebra. The `classical' limit $z\to 0$ (or, equivalently, $q\to 1$) provides the corresponding Lie--Poisson coalgebra. In fact, these $q$-Poisson coalgebras can be written as quadratic Poisson structures on certain (dual) Lie--Poisson groups, that are invariant with respect to the group multiplication (which is nothing but a coproduct map on the dual, see~\cite{BRcluster}).

\end{itemize}

Some remarks are in order.

\begin{itemize} 

\item In general, we cannot say {\em a priori} that $H^{(N)}$ is a completely integrable Hamiltonian system because~\cite{alfonso}:
(i)  We have to determine the  number of  degrees of freedom of $H^{(N)}$ by choosing a explicit symplectic realization of the coalgebra $(A,\Delta)$. (ii)   Once this number is fixed, we have to check whether the  number of independent invariants extracted from $C^{(m)}_j$ (under such a specific symplectic realization) is enough to guarantee complete integrability.

\item However,  if the coalgebra symmetry provides the complete integrability for $H^{(N)}$,   the integrals of the motion can always be  obtained in arbitrary dimension $N$ in an explicit form.  Hence the coalgebra symmetry arises as a unified approach to integrability since we obtain  families of $N$D systems that share a very large common set of integrals of the motion. Furthermore,  in this case $H^{(N)}$   will probably be    {superintegrable} (to some extent)~\cite{CRMAngel}.

\item   In the case of $q$-deformations of Poisson coalgebras, the Hamiltonians obtained through the coalgebra approach are  {integrable deformations} of the $z=0$  ($q=1$) cases in such a manner that different quantum algebras give rise to different integrable deformations.   Moreover, this fact can provide some relevant information concerning the {geometric/physical interpretation} of the deformation parameter.

\item   This construction  {holds for noncommutative coalgebras} aswell. Thus,  quantum mechanical systems can also  be constructed (although ordering problems have to be fixed).

\end{itemize}

Before entering into explicit examples of the coalgebra construction, two important and general aspects have to be more deeply analysed in order to provide a global overview on the subject: they are the role of the symplectic realizations (which has been thoroughly studied in~\cite{alfonso}) and the
superintegrability features of the coalgebra symmetry (which were presented for the first time in~\cite{CRMAngel}). 


\subsection{Symplectic realizations and complete integrability} \label{Symplectic}

Let  $(A,\Delta)$  a Poisson coalgebra  with  {$r$ Casimir
functions} ${\cal C}_j$ $(j=1,\dots, r)$. A {symplectic leaf} of $A$ (which is always even-dimensional) 
will be denoted by
$A_{(k_1,k_2,\dots, k_r)}$, where the leaf is characterized by 
a given set of constant values $(k_1,k_2,\dots, k_r)$ for the  Casimirs.

An {\em $s$-dimensional symplectic realization}  $D$ for $A_{(k_1,k_2,\dots,
k_r)}$ is given (locally) in terms of
$s$ pairs $(q_i,p_i)$ of canonical Darboux variables
\be
D:x\rightarrow 
x(q_1,p_1,q_2,p_2,\dots,q_s,p_s) 
\label{za}
\ee
where $x$ is any point on $A_{(k_1,k_2,\dots,k_r)}$. We remark that, in principle, 
 {different symplectic leaves} $A_{(k_{1,i},k_{2,i},\dots,k_{r,i})}$ can be chosen  {for each copy $i$ of $A$} within  $A\otimes
A\otimes\dots^{N)}\otimes A$.

In particular, if we consider symplectic realizations with the
same $s$ for all the $N$ sites in the tensor product $A\otimes
A\otimes\dots^{N)}\otimes A$, the Hamiltonian (\ref{htotg})
\be
H^{(N)} =(D\otimes D\dots^{N)}\otimes
D)(\co^{(N)}({\cal H}))
\ee 
 turns out
to be a function of $N\cdot s$ pairs of canonical variables, i.e., it defines
 {a system with
$N\cdot s$ degrees of freedom}. Then for  {each nonlinear Casimir} ${\cal{C}}_j$, we get at most $(N-1)$ integrals coming from its $m$-th coproducts  (\ref{Ctotg}) written under the  $m$-th  tensor product of the symplectic realization $D$ (\ref{za}):  
\be
C^{(m)}_j =(D\otimes D\dots^{m)}\otimes
D)(\co^{(m)}({\cal{C}}_j))
\ee 
where $m=2,\dots,N$;  notice that $C^{(1)}_j=D({\cal C}_j)=k_j$.
Note that under symplectic realizations  $m$-th coproducts of  {linear Casimirs  always  give just numerical constants}.

 Thus, if we have  {$R$ nonlinear Casimirs} we find a maximum possible number of integrals in involution given by
$$
(N-1)\cdot R .
$$
In order to get complete integrability, we should have that
$$N\cdot s - 1\leq (N-1)\cdot R$$ thus we need that the chosen symplectic
realization
$D$ of
$A$ fulfils (for any $N$)
\be
 {
s\leq R-\frac{R-1}{N} .
}
\label{conda}
\ee
Consequently, 
  the  {\em necessary condition for complete integrability}  connects the dimension of the symplectic realization and the number of nonlinear Casimir functions through the conditions~\cite{alfonso}:
  \begin{itemize}

\item { $s=1$ } for coalgebras with  { {\bf $R=1$}}.

\item { $s < R$ } for coalgebras with   { {\bf $R>1$}}.

\end{itemize}

 Let us now  consider a particular type of symplectic realizations whose dimension is  fixed by the dimension $l$   of the Poisson coalgebra and its number $r$ of Casimir functions. We shall call `generic' to the symplectic realization with maximal dimension $s_m$
given by
 {\be
s_m=\frac{l-r}{2} .
\label{condb}
\ee}
 Hence, the  {\em integrability  condition} for the generic symplectic realization is
$$
s_m=\frac{l-r}{2}\leq R-\frac{R-1}{N}
$$
which leads to the final expression
 {\be
 {
l\leq (2\,R + r)-\frac{2}{N}(R-1) .
}
\ee}

 Therefore, complete integrability for the generic symplectic realization can be  achie\-ved if~\cite{alfonso}:

\begin{itemize}

\item {$l\leq  2 + r$} for coalgebras with    {$R=1$}.

\item {$l< 2\,R + r$} for coalgebras with     {$R>1$}.

\end{itemize}

As a byproduct of the above results, if we consider  {\em simple Lie algebras}, with rank $r=R$,   we find  that for  coalgebras with {$R=1$}, their dimension must be $l\leq 3$, and for coalgebras with  {$R>1$},  then we get the condition $l< 3R$. This, in turn, shows that only simple Lie algebras of rank 1 can provide complete integrable systems in the generic symplectic realization, whilst all higher-rank ones do not fulfill the condition.

Another family of interesting cases is given by  Lie coalgebras in which the generic symplectic realization has $s_m=1$, since  the coalgebra symmetry always  satisfies the integrability condition provided  that $R\geq 1$. This possibility covers many {\em non-simple Lie algebras} fulfilling $l-r=2$.


\subsection{Coalgebra superintegrability from coassociativity}

Instead of  (\ref{fl}), another recursion relation for
the {$m$-th coproduct} map can be defined as~\cite{CRMAngel}: 
\be
\co_R^{(m)}:=(\co^{(2)}\otimes {\rm id}\otimes\dots^{m-2)}\otimes
{\rm id}\,)\circ\co_R^{(m-1)}.
\label{zb}
 \ee
Due to the coassociativity property of the coproduct, this new expression 
will provide exactly the same expressions for the $N$-th coproduct of
any generator of $A$: 
\be
\Delta^{(N)}(X_i)\equiv \Delta_R^{(N)}(X_i).
\label{zc}
\ee
However, if we label from
$1$ to $N$ the sites of the chain $A\otimes
A\otimes\dots^{N)}\otimes A$, lower dimensional
$m$th-coproducts
 (with $2<m<N$) will be `different' in the sense that the `left' coproducts
$\co^{(m)}$  (\ref{fl})  will contain objects living on the tensor product space
$ 1\otimes 2\otimes \dots\otimes m $, whilst  the `right' coproducts $\co_R^{(m)}$ will be
defined on the sites
$(N-m+1)\otimes (N-m+2)\otimes \dots\otimes N$.

Therefore, the coalgebra symmetry of a given
Hamiltonian  gives rise to two `pyramidal' sets of $r\cdot N$ integrals of
the motion in involution that Poisson-commute with
$H^{(N)}$~\cite{CRMAngel}, under  the corresponding symplectic realization $D$ (\ref{za}). Namely, both sets are just the `left' integrals above considered $C^{(m)}_{j}$ (\ref{Ctotg}) together with the `right' ones given by
\be
C_{j,{(m)}}:= \co_R^{(m)}({\cal{C}}_j(X_1,\dots,X_l))=
{\cal{C}}_j(\co_R^{(m)}(X_1),\dots,\co_R^{(m)}(X_l)) 
\label{Ctotgb}
\ee
such that $C^{(N)}_{j}\equiv C_{j,{(N)}}$, while both $C^{(1)}_{j}$ and  $C_{j,{(1)}}$ are  constants.

The very same arguments discussed in section 2.1 on the connection between integrability and the dimensionality of the chosen symplectic realizations can be applied to the {\em additional} set of `right' integrals $C_{j,{(m)}}$. In this way, a completely integrable Hamiltonian $H^{(N)}$ with coalgebra symmetry will be, in principle, {\em superintegrable}.

For the sake of symplicity let us assume that we are dealing with a Poisson coalgebra  with  $R=1$, that is, with $N$ degrees of freedom and with a single nonlinear Casimir. Therefore if we take a symplectic realization with $s=1$  the necessary condition for integrability (\ref{conda}) is fulfilled.
In this case, the coalgebra approach gives rise to  $(2N-3)$ functionally independent integrals, which are displayed in table~\ref{table2}, and with  $\Delta^{(N)}({\cal{C}})\equiv
\co_R^{(N)}({\cal{C}}) $ being a common `left-right' integral. Then each set of `left' and `right' integrals is then formed by $(N-1)$ functions in involution. Therefore 
  there is  {\em only one} missing integral  in order  to ensure the {\em maximal 
superintegrability} of the system, which means that $H^{(N)}$ is always,  at least,  QMS. Such a `remaining' integral, in case it does exist,  will not be provided by the coalgebra symmetry and it will have to be found by alternative methods.
Hence, we can conclude that the coalgebra symmetry for this class of Hamiltonian systems with $N$ degrees of freedom implies the following integrability hierarchy:

\begin{itemize}
 
\item  {$N=2$}. The Hamiltonian is only {\em integrable} with a single constant of motion $C^{(2)}= C_{(2)}$.

\item {$N=3$}. The Hamiltonian  is   {\em minimally} (or weakly) superintegrable with three integrals given by $\{C^{(2)},C^{(3)}= C_{(3)},C_{(2)}\}$.

\item   {$N>3$}. The Hamiltonian is   {\em QMS}  with $(2N-3)$ integrals $\{C^{(m)},C^{(N)}= C_{(N)},C_{(m)}\}$ for $m=2,\dots,N-1$.

\end{itemize}

\begin{table}[t]
{\footnotesize
 \noindent
\caption{{Coalgebra symmetry and QMS: the $(2N-3)$ integrals   coming from a   Casimir ${\cal C}$.}}
\label{table2}
\medskip
\noindent\hfill
$$
\begin{array}{cl}
\hline
&\\[-6pt]
\multicolumn{1}{c}\rm{`Left' set of $(N-1)$ integrals $C^{(m)}$ in involution}\quad&  \mbox{Tensor product space}\\[4pt]
 C^{(2)}    \equiv\Delta^{(2)}({\cal C}) 
&   1\otimes 2 \cr  
C^{(3)} \equiv\Delta^{(3)}({\cal
C})&   1\otimes 2\otimes 3\cr 
   \vdots    &\qquad \vdots \cr  C^{(m)} 
 \equiv\Delta^{(m)}({\cal C})   & 1\otimes 2\otimes
\dots
\otimes m\cr
   \vdots    &\qquad \vdots \cr  
 C^{(N)} \equiv\Delta^{(N)}({\cal C}) &  1\otimes 2\otimes
\dots \otimes (N-1)
\otimes N \\[6pt]
\multicolumn{1}{c}\rm{`Right' set of $(N-1)$ integrals $C^{(m)}$ in involution}\quad&  \mbox{Tensor product space}\\[4pt]
  C_{(2)}  \equiv\Delta^{(2)}_R({\cal C})   & (N-1)\otimes N \cr 
 C_{(3)}   \equiv\Delta^{(3)}_R({\cal C})  & (N-2)\otimes (N-1)
\otimes N\cr
  \vdots    &\qquad \vdots \cr  
   C_{(m)}  \equiv\Delta^{(m)}_R({\cal C}) &
(N-m+1)\otimes (N-m+2)\otimes \dots\otimes N \cr
   \vdots    &\qquad \vdots \cr  
  C_{(N)}=C^{(N)}
\equiv\Delta^{(N)}_R({\cal C})  &
1\otimes 2\otimes 
\dots
\otimes (N-1)\otimes N\\[4pt]
 \hline
 \end{array}
$$
\hfill}
\end{table}


\section{Three Poisson coalgebras} \label{Three Poisson coalgebras}

We stress that the coalgebra approach to complete integrability  is completely general and constructive for  any Poisson coalgebra endowed
with a {\em suitable} symplectic realization. In order to illustrate the above ideas, we will mainly  consider in this contribution the following three Poisson coalgebras: (i) $\frak{sl}(2,\R)$,  (ii)   its non-standard $q$-deformation $\frak{sl}_z(2,\R)$ and (iii) the (two-photon) algebra $h_6$. With them 
a bunch of physically interesting $N$D Hamiltonians can be obtained, and some of them will be explicitly presented in sections 4,  5 and 6, respectively.


\subsection{The $\frak{sl}(2,\R)$ Lie--Poisson coalgebra}

  This coalgebra is defined by the following Lie--Poisson brackets and comulti\-plication map: 
\begin{equation}
 \{J_3,J_+\}=2 J_+     \qquad  
\{J_3,J_-\}=-2 J_- \qquad   
\{J_-,J_+\}=4 J_3    
\label{ba}
\end{equation}
\begin{equation}
\begin{array}{l}
\Delta(1)=1\otimes 1\qquad \Delta(J_i)=  J_i \otimes 1+ 1\otimes J_i \qquad i=+,-,3.
\end{array}
\label{bb}
\end{equation}
The   Casimir function for $\frak{sl}(2,\R)$ reads
\begin{equation} 
{\cal C}=  J_- J_+ -J_3^2  . 
\label{bc}
\end{equation} 
A one-particle symplectic realization of this coalgebra is given by 
\be
D(\jm)=q_1^2 \qquad
D(\jp)=p_1^2 +
\frac{\otra_1}{q_1^2} \qquad
D(\jj)= q_1 p_1  ,
\label{bbcc}
\ee
where $\pois{q_1}{p_1}=1$. Note that, under this realization, $
C^{(1)}=D({\cal C})=b_1$. Hence, according to the notation and results presented in section 2.1 we are dealing with a Poisson coalgebra with dimension $l=3$ and with a single nonlinear Casimir. Thus $r=R=1$ and this implies that (\ref{bbcc}) is the generic  symplectic realization with $s\equiv s_m=(l-r)/2=1$.   The  corresponding $N$-particle symplectic realization of $\frak{sl}(2,\R)$,  
living on  
$\frak{sl}(2,\R)\otimes  \dots^{N)}\otimes \frak{sl}(2,\R) $, is obtained by applying
$$
J_i^{(N)}=(D\otimes D\otimes\dots^{N)}\otimes D)(\Delta^{(N)}({J_i}))
$$
which gives~\cite{Deform,BHletter} 
\bea
 &&J_-^{(N)}=\sum_{i=1}^N q_i^2\equiv \>q^2 \qquad    J_3^{(N)}=
  \sum_{i=1}^N  q_i p_i\equiv \>q\cdot\>p  \cr
 &&  J_+^{(N)}=
    \sum_{i=1}^N \left(  p_i^2+\frac{\otra_i}{ q_i^2} \right)
\equiv \>p^2 +  \sum_{i=1}^N  \frac{\otra_i}{ q_i^2} ,
\label{be}
\eea
where $\otra_i$ are $N$ arbitrary real parameters. This means that  the $N$-particle
generators (\ref{be}) fulfil the  commutation rules (\ref{ba}) with respect to the  
$N$D canonical Poisson bracket. 

Next, since $s_m=1$ and $R=1$ we are in the case of a Poisson coalgebra endowed with 
  the $(2N-3)$ integrals displayed in table 2; these turn out to be~\cite{CRMAngel,BHletter}:
\be
  C^{(m)}= \sum_{1\leq i<j}^m  I_{ij}
+\sum_{i=1}^m \otra_i \qquad 
 C_{(m)}= \sum_{N-m+1\leq i<j}^N I_{ij}
+\sum_{i=N-m+1}^N \otra_i 
\label{cinfm}
\ee
where  $m=2,\dots,N$ and
\be
I_{ij}=({q_i}{p_j} - {q_j}{p_i})^2 + \left(
\otra_i\frac{q_j^2}{q_i^2}+\otra_j\frac{q_i^2}{q_j^2}\right)
\ee
are the $b_i$-generalization of the square of the  `angular momentum' generators $J_{ij}={q_i}{p_j} - {q_j}{p_i}$ which span an $\frak{so}(N)$ Lie--Poisson algebra.  As a consequence of the coalgebra symmetry, the generators (\ref{be})
Poisson commute with these
$(2N-3)$ functions. Therefore, any   arbitrary function ${\cal H}$ defined as
\be
{H}^{(N)}= {\cal H}\left(J_-^{(N)},J_+^{(N)},J_3^{(N)}\right)
={\cal H}\left(\>q^2,\>p^2+\sum_{i=1}^N\frac{\otra_i}{q_i^2},\>q\cdot\>p  \right)
\label{classham}
\ee
gives rise to an $N$D QMS Hamiltonian system which is always endowed, at least, with the $(2N-3)$ integrals (\ref{cinfm}).


\subsection{$q$-Poisson coalgebras: the $\mathfrak{sl}_z(2,\R)$ case}

The Poisson analogues of quantum
algebras and groups~\cite{Dri,CP} are also (deformed) coalgebras
$(A_z,\Delta_z)$ ($q={\rm e}^z$), which means that 
any function of the generators of a given `quantum' Poisson algebra (with
deformed Casimir elements ${\cal C}_{z,j}$) will provide a deformation of the 
Hamiltonian
generated by the undeformed coalgebra. Such a $q$-coalgebraic  deformation will preserve, by
construction, the (super)integrability properties of the system defined on the undeformed Lie--Poisson coalgebra.
Therefore,
$q$-deformations can be understood in this context as the algebraic machinery suitable for generating integrable
deformations of Hamiltonian systems.

In particular, let us focus on the non-standard  
$\mathfrak {sl}_z(2,\R)$ Poisson coalgebra   defined by the following
(deformed) Poisson brackets and  coproduct map (see~\cite{Deform,Ohn}):
\begin{equation}
\{\jj,\jp\}=2 \jp \cosh z\jm  \qquad
 \{\jj,\jm\}=-2\,\frac {\sinh z\jm}{z} \qquad
 \{\jm,\jp\}=4 \jj 
\label{baa}
\end{equation}
\bea
&&\Delta_z(1)=1\otimes 1\qquad  \Delta_z(\jm)=  \jm \otimes 1+
1 \otimes \jm \nonumber\\
&&  \Delta_z(J_i)=J_i \otimes {\rm e}^{z \jm} + {\rm e}^{-z \jm} \otimes
J_i  \qquad (i=+,3).
\label{ggcc}
\eea
The Casimir function for $\mathfrak {sl}_z(2,\R)$ reads
\begin{equation}
\qquad\qquad{\cal C}_z=  \frac {\sinh z\jm}{z}\, \jp -\jj^2 .
\label{gc}
\end{equation}
A one-particle (deformed) symplectic realization of $\mathfrak {sl}_z(2,\R)$  is:
\be
D_z(\jm)=q_1^2\qquad
 D_z(\jp)=\frac {\sinh z q_1^2}{z q_1^2}\, p_1^2 +
\frac{z \otra_1}{\sinh z q_1^2} \qquad
 D_z(\jj)=\frac {\sinh z q_1^2}{z q_1^2}\, q_1 p_1  
 \label{zaa}
\ee
such that $C_z^{(1)}=D_z({\cal
C}_z)=\otra_1$. Hence we are dealing again with a coalgebra with $l=3$, $r=R=1$ and $s_m=1$. The $N$th-coproduct of (\ref{ggcc}) through  the one-particle  representation (\ref{zaa}) gives rise to an  
$N$-particle symplectic realization on $ \mathfrak {sl}_z(2,\R) \otimes  \dots^{N)}\otimes \mathfrak {sl}_z(2,\R) $ through
$$
J_i^{(N)}=(D_z\otimes D_z\otimes\dots^{N)}\otimes D_z)(\Delta_z^{(N)}({J_i})),
$$
namely
\bea  
&& 
\jm^{(N)}= \sum_{i=1}^N q_i^2 \equiv \>q^2  \nonumber \\ 
&& \jp^{(N)}=\sum_{i=1}^N
\left( \frac {\sinh z q_i^2}{z q_i^2} \, p_i^2  +\frac{z \otra_i}{\sinh z
q_i^2} \right)  \exp\left\{ -
z\sum_{k=1}^{i-1}  q^2_k+ z
\sum_{l=i+1}^N q^2_l   \right\}       \equiv \tilde{\>p}_z^2  \nonumber \\ 
&&\jj^{(N)}=\sum_{i=1}^N
\frac {\sinh z q_i^2}{z q_i^2} \, q_ip_i \exp\left\{ -
z\sum_{k=1}^{i-1}  q^2_k+ z
\sum_{l=i+1}^N q^2_l   \right\}      
\equiv (\>q\cdot\>p)_z
\label{zsymp}
\eea
where the  $\otra_i$'s are again  $N$ arbitrary real
parameters that label the representation on each `lattice' site.

In this case the $(2N-3)$ functions written in table 2 are expliciltly given by~\cite{Deform, CRMAngel}:
\bea
&&
C_z^{(m)}= \sum_{1\le i<j}^m{I_{ij}^z}\,\exp\left\{   - 2 z \sum_{k=1}^{i-1}   q^2_k  -    z  q^2_i  +   z  q^2_j  + 
2 z \sum_{l=j+1}^m   q^2_l  \right\} \nonumber\\
&&\qquad\qquad  +\sum_{i=1}^m \otra_i  \exp\left\{
   -2 z
\sum_{k=1}^{i-1}  q^2_k+ 2 z
\sum_{l=i+1}^m   q^2_l    \right\} \nonumber\\
&& C_{z,{(m)}} =  \sum_{N-m+1\leq i<j}^N\!\!\!\!  I_{ij}^z \exp\left\{   - 2 z\!\!  \sum_{k=N-m+1}^{i-1}  \!\! 
q^2_k  -  z   q^2_i  +    z q^2_j  +  2z \sum_{l=j+1}^N   q^2_l   \right\} \nonumber\\
&&\qquad\qquad 
+\sum_{i=N-m+1}^N \!\! \!\!  \otra_i  \exp\left\{
- 2 z \!\!  \sum_{k=N-m+1}^{i-1}  \!\!  q^2_k+ 2 z
\sum_{l=i+1}^N   q^2_l    \right\} 
\label{zcc}
\eea
 where  $m=2,\dots,N$ and
 \be
I_{ij}^z= \frac {\sinh z q_i^2}{z  q_i^2}\,
\frac {\sinh z q_j^2}{z q_j^2}
\left({q_i}{p_j} - {q_j}{p_i}\right)^2  
 +\left( \otra_i\, \frac {\sinh z q_j^2}{\sinh z q_i^2}
+ \otra_j \,\frac {\sinh z q_i^2}{\sinh z q_j^2} \right) .
\ee

Consequently, any   smooth
function 
  ${\cal H}_z$ defined on the
$N$-particle symplectic realization (\ref{zsymp}) of the generators of  $\mathfrak {sl}_z(2,\R)$ in the form
\begin{equation}
{H}^{(N)}_z= {\cal H}_z\left(\jm^{(N)},\jp^{(N)},\jj^{(N)}\right)
={\cal
H}_z\left(\>q^2,\tilde{\>p}_z^2,(\>q\cdot\>p)_z
\right) 
\label{hamham}
\end{equation}
defines a QMS Hamiltonian system. We stress that all the choices for ${\cal H}_z$ share the `universal' set of $(2N-3)$ constants of motion given by (\ref{zcc}).

Clearly, the non-deformed limit $z\to 0$ of all the above expressions gives rise to the ones  corresponding to the  $\mathfrak {sl}(2,\R)$-coalgebra presented in section 3.1, which shows that quantum algebras may provide (super)integrable generalizations of non-deformed Hamiltonian systems.

 We also  remark that this quantum deformation has been interpreted in~\cite{Deform} as an algebraic way to introduce  long-range interactions in the underlying underformed systems due to  the   exponentials of the type $\exp(z \sum_j q_j^2)$  coming from the deformed symplectic realization  (\ref{zsymp})  (compare with (\ref{be}) and (\ref{classham})).


\subsection{The two-photon Lie--Poisson coalgebra}

The third coalgebra that we explicitly consider  also satisfies  the {\em necessary condition} (\ref{conda}) for complete integrability. Nevertheless, we shall use this example to show that  the latter   is {\em not}  a {\em sufficient condition}, thus making evident the essential role played by the symplectic realization of the chosen coalgebra.

The two-photon coalgebra $(h_6,\Delta)$ is spanned   by six   generators
$\{\kk,A_+,A_-,B_+,\\B_-,M\}$ with  Lie--Poisson brackets given by~\cite{BHnonlin}
\be 
\begin{array}{lll}
\{\kk,A_+\}=A_+&\qquad \{\kk,A_-\}=-A_- &\qquad \{A_-,A_+\}=M\cr
\{\kk,B_+\}=2B_+&\qquad \{\kk,B_-\}=-2 B_- &\qquad \{B_-,B_+\}=4\kk+2M\cr
\{A_+,B_-\}=-2 A_-  & \qquad \{A_+,B_+\}=0& \qquad
\{M,\,\cdot\,\}=0\cr
\{A_-,B_+\}=2A_+&\qquad \{A_-,B_-\}= 0& 
\end{array}
\label{aa}
\ee
together with the usual nondeformed coproduct map given by (\ref{standardcop}). This is just the Poisson analogue of the (non-simple)  $h_6$ Lie algebra which has been widely applied in quantum optics~\cite{Gil, Brif}. The $h_6$ algebra is isomorphic to the $(1+1)$D  Schr\"odinger algebra~\cite{twop} and the following embedding of the Heisenberg--Weyl $h_3$, harmonic oscillator $h_4$ and  two-photon coalgebras can be easily identified:
$$
(h_3,\Delta)\subset (h_4,\Delta)\subset (h_6,\Delta)\qquad  h_3=\langle A_-,A_+,M\rangle\qquad h_4=\langle \kk,A_-,A_+,M\rangle.
$$ 

The $h_6$-coalgebra is endowed with two Casimir operators~\cite{Patera}, namely
\be
{\cal C}_1=M\qquad {\cal C}_2= (MB_+ -A_+^2)(MB_- - A_-^2)-
(M \kk -   A_- A_+ +M^2/2)^2  .
\label{aac}
\ee

Let us now  introduce   the one-particle symplectic realization of $h_6$ given by~\cite{2photon,BHnonlin}
\be 
\begin{array}{lll}
D(A_+)=\la_1  p_1&\quad
D(A_-)=\la_1  q_1 &\quad   
D(\kk)=q_1\,  p_1  -\frac 12  {\la_1^2 }\cr
\displaystyle{ D(B_+)= {p_1^2}}
&\quad D(B_-)= 
q_1^2   &\quad   D(M)=\la_1^2 
\end{array}
\label{ad}
\ee 
where $\lambda_1$ is a non-vanishing constant that labels the   realization such that  $
C^{(1)}_1=D({\cal C}_1)=\la_1^2$ and $
C^{(1)}_2=D({\cal C}_2)=0$. Notice that if $\la_1=0$ the $h_6$-coalgebra reduces to the $\mathfrak {sl}(2,\R)=\langle B_-,B_+,\kk\rangle$ one (\ref{bbcc}) with $b_1=0$.  In fact, for any $\la_1\ne 0$,  $\mathfrak {gl}(2,\R)=\langle B_-,B_+,\kk,M\rangle$ is also a sub-coalgebra of $h_6$.

Therefore, by taking into account section 2.1 we are now  dealing with a coalgebra with $l=6$, $r=2$, $R=1$ $({\cal C}_1=M$ is linear and leads to trivial integrals) and the chosen symplectic realization has $s=1$. This means that the necessary condition for integrability (\ref{conda}) is fulfilled since 
$s=1\equiv R= 1 $, although note that the symplectic realization is not a generic one (\ref{condb})  since    $s_m=1\ne \frac{l-r}{2}=2$. 

By making use of the coproduct map it is immediate to show that the    $N$-particle symplectic realization on 
$h_6 \otimes  \dots^{N)}\otimes h_6$ reads
\bea
&& A_+^{(N)} =\sum_{i=1}^N \la_i p_i   \equiv \bla\cdot \>p \qquad
 A_-^{(N)} =
\sum_{i=1}^N \la_i  q_i   \equiv \bla \cdot \>q  \qquad   M^{(N)} =\sum_{i=1}^N
\la_i^2  \equiv\bla^2\cr 
&& B_+^{(N)}=\sum_{i=1}^N 
{p_i^2}\equiv \>p^2 \quad\
 B_-^{(N)} = 
\sum_{i=1}^N  q_i^2\equiv \>q^2\quad\
   \kk = \sum_{i=1}^N
\bigg(q_i p_i -\frac {\la_i^2}2\bigg) \equiv \>q\cdot\>p -\frac  { \bla^2}2 \nonumber\\
&&
\label{sympn}
\eea 
where $\bla=(\la_1,\dots,\la_N)$.  From this viewpoint it is clear that  the $h_6$-coalgebra  may give  rise to a generalization of the integrable systems with $\mathfrak {sl}(2,\R)$-coalgebra symmetry, provided that  {\em all} $b_i=0$ but, as we shall show in what follows, one has to pay, in principle,   the price of loosing {\em complete integrability}.
We   stress that the $h_6$ Poisson algebra was formerly considered in~\cite{Rauchh6} in the framework of  some  3D integrable systems, but only the additional coalgebra structure here presented makes possible to consider $h_6$ as a useful symmetry in $N$D systems.  

We can now compute the $(2N-3)$ integrals of motion displayed in table 2 and coming from the nonlinear Casimir
\be
{\cal C} ={\cal C} _2 /{{\cal C} _1}
 =MB_+B_- - B_+ A_-^2  - B_- A_+^2  - M(\kk+M/2)^2 + 2
A_-A_+(\kk+M/2)  .
\label{ac}
\ee 
A straightforward computation shows that  $C^{(2)}=C_{(2)}=0$ and the remaining ones read
\bea
&& C^{(m)}=
{\sum_{1\leq i<j<k}^{m}} 
\left(
\la_i (p_j q_k - p_k q_j ) +
\la_j (p_k q_i - p_i q_k ) +
\la_k (p_i q_j - p_j q_i ) 
\right)^2\nonumber\\
&&C_{(m)}=
{\sum_{ N-m+1\le i<j<k}^{N}} 
\bigl(
\la_i (p_j q_k - p_k q_j ) +
\la_j (p_k q_i - p_i q_k ) +
\la_k (p_i q_j - p_j q_i ) 
\bigr)^2 
\label{c3m}
\eea
where $m=3,\dots, N$. This means that   any smooth function 
\bea
&&{H}^{(N)}={\cal H}\left(\kk^{(N)} ,B_-^{(N)} ,B_+^{(N)} ,A_-^{(N)} ,A_+^{(N)} ,M^{(N)} \right)\nonumber\\
&&\qquad\,  =
{\cal H}\left(\>q\cdot \>p-   \frac12 \bla^2    ,\>q^2,\>p^2 ,\bla\cdot\>q,   \bla\cdot\>p ,  \bla^2 \right)
\label{classhamb}
\eea
Poisson commutes with the $2N-5$ functions (\ref{c3m}) but now the sets $\{C^{(3)},\dots, C^{(N)} \}$ and $\{C_{(3)},\dots, C_{(N)} \}$ are only formed by $N-2$ integrals in involution so that, in general,  ${H}^{(N)} $ does {\em not} provide a completely integrable system. Since there is only one constant left, say $\cal I$, to obtain complete integrability we have called these systems {\em quasi-integrable} ones~\cite{2photon}. Nevertheless, this initial failure of the coalgebra approach can be circumvented by making use of the rich subalgebra structure of $h_6$ and, for some particular choices of $ {H}^{(N)}$, such a remaining integral can be obtained as follows (see~\cite{2photon} for a complete discussion on the subject). 

Firstly, in order to ensure the existence of ${\cal I}$ for any dimension $N$, we shall assume that this additional integral is also $h_6$-coalgebra invariant, which means that it can be written as a function
\be
{\cal I}={\cal I}(\kk ,B_+ ,B_- ,A_+ ,A_- ,M )
\label{Iinv}
\ee
where the $h_6$ generators are written in their $N$-particle symplectic realization (\ref{sympn}). In this way, if  ${\cal I}$ is functionally independent with respect to both the $h_6$ Casimir (\ref{ac}) and the Hamiltonian ${\cal H}$, 
the coalgebra symmetry guarantees --by construction-- the involutivity of ${\cal I}$ with respect to the $(N-2)$ `left' integrals $C^{(m)}$ $(m=3,\dots,N)$ and its  functional independence with respect to them. And the very same result holds for the 
  $(N-3)$  `right'  integrals $C_{(m)}$ where $m=3,\dots,N-1$ (we recall that $C_{(N)}=C^{(N)}$).  Therefore, if $\cal I$ can be finally  found then the corresponding  Hamiltonian is not only integrable but furthermore superintegrable with $2N-4$ functionally independent constants of motion (one less than a QMS system).

In particular, we can consider two different situations in which the existence of ${\cal I}$ is guaranteed  by construction:

\medskip

\noindent $\bullet$ {\em Subalgebra integrability}.  If the Hamiltonian ${\cal H}$ is defined within a subalgebra of $h_6$ that has a nonlinear Casimir invariant, the $N$-particle realization of the Casimir of the subalgebra provides automatically the integral ${\cal I}$.

 \medskip

\noindent $\bullet$ {\em Generator integrability}.  Now, let us choose a given  generator $X$ of $h_6$. If we look for all the generators $X_{j}\,\,\,\,  ( j=1,\dots, n$) commuting with $X$ and we also look  for all the subalgebras $g_{k}\,\,\,(k=1,\dots, t$) containing $X$ as generator, then the Hamiltonian constructed through any function of the type
\be
{\cal H}_{X}={\cal H}_X\left( {\cal C}_{g_{1}},\dots,  {\cal C}_{g_{t}}, X,X_{1},\dots, X_{n}\right),
\label{hx}
\ee
where $ {\cal C}_{g_{k}}$ is the Casimir function of the subalgebra ${g_{k}}$, is such that
\be
\{{\cal H}_{X},X\}=0.
\ee
Moreover, the $N$-th particle symplectic realization of both $X$ and ${\cal H}_{X}$  will Poisson-commute with the two sets of  integrals $ C^{(m)}$  and    $C_{(m)}$ (\ref{c3m}). Under such hypotheses, ${\cal H}_X$ is completely integrable since the $N$-th particle symplectic realization of the generator $X$ is just the additional constant of   motion ${\cal I}$. Since we have five relevant generators $\{\kk ,B_+ ,B_- ,A_+ ,A_- \}$  this procedure will give rise to five families of completely integrable systems that have been  studied in detail in~\cite{2photon}. As we shall see in section 6, $h_6$-coalgebra systems include natural Hamiltonians as well as static electromagnetic fields and curved geodesic flows. 

But even in the case that a given Hamiltonian does not fit within the two previous cases, the search for the remaining integral ${\cal I}$ --in case it does exist-- can  also  be performed by using direct methods that can be computerized. In fact,  the additional integral $\mathcal{I}$ can be searched among $h_6$ functions with cubic or higher dependence on the momenta (note that all the integrals that we have presented so far are, at most, quadratic in the momenta). Indeed, some particular solutions to this problem have recently  been found  and lead to new completely integrable $N$D Hamiltonians (see~\cite{2photon,perturb}).

Finally, it could also happen that for a certain ${\cal H}$ the additional integral ${\cal I}$ does exist, but  it cannot be written as a function (\ref{Iinv}) of the $h_6$ generators ({\em i.e.}, ${\cal I}$ is not coalgebra-invariant). This implies that  the explicit form for ${\cal I}$ has to be found for each dimension, and --hopefully-- a generic $N$D expression can be found inductively. Nevertheless, due to the physical interest of $h_6$ invariant systems (see section 6) this possiblity is worth to be explored with the aid of symbolic computation packages.


\section{QMS Hamiltonians with $\frak{sl}(2,\R)$-coalgebra\\ symmetry}

In the sequel we will show the potentialities of the coalgebra framework by summarizing some of the $\frak{sl}(2,\R)$-coalgebra symmetric Hamiltonians of the form  (\ref{classham}) that have been so far described in the literature. Some of them were already known but many others have been
constructed for the first time by making use of the constructive approach detailed in  
section 3.1. Since we are always dealing with $N$D systems  and  for the sake of simplicity of the notation, from now on we shall drop the index `$(N)$' in both the $N$D symplectic realization and the Hamiltonians.


\subsection{Evans systems}

The following generalization of the motion of a classical particle on an $N$D Euclidean space $\mathbb E^N$
under a spherically symmetric potential is   $\frak{sl}(2,\R)$-coalgebra invariant~\cite{Deform}
\be 
 H =\frac 12 \jp +{\cal F} \left( \jm \right)= \frac 12 \>p^2 +
{\cal F}\left(\>q^2\right)+  \sum_{i=1}^N   \frac{\otra_i}{2q_i^2} 
\label{anha}
\ee
where ${\cal F}$ is a   smooth function and the terms depending on the $\otra_i$'s  constants are just additional `centrifugal barriers' coming from a non-zero symplectic realization of  $\frak{sl}(2,\R)$. In fact, the  $\frak{sl}(2,\R)$-coalgebra provides a common set of (right and left) integrals of the motion (\ref{cinfm}) for any choice
of ${\cal F}$~\cite{BHletter}. Note that the $N$D Smorodinsky--Winternitz  Hamiltonian~\cite{Fris} is
recovered when 
${\cal F}(\>q^2) =\omega^2\>q^2/2$, and the Kepler--Coulomb
potential corresponds to ${\cal F}(\>q^2) =-k/|\>q|$, where $\omega$ and $k$
are real constants (and $|\>q|= \sqrt{\>q^2}$).
Therefore, both outstanding MS systems do have coalgebra symmetry.


\subsection{Superintegrable electromagnetic field Hamiltonians}
Certain velocity-dependent potentials on $\mathbb E^N$   giving rise to superintegrable
electromagnetic field Hamiltonians~\cite{Dorizzi,winter}  can also  be obtained in this framework.  The most
general example of this class is given by~\cite{BHletter}:
\bea
&& H  = \frac 1{2 }\jp   - e   \jj 
 \funciona\left (\jm \right)+ e \funcionb\left(\jm \right)    \nonumber\\
&&\qquad\, =\frac 1{2 }\, \>p^2-  e  
(\>q\cdot\>p) \funciona(\>q^2)+ e \funcionb(\>q^2)  +\sum_{i=1}^N \frac{\otra_i}{ 2  
q_i^2} 
 \label{cf}
\eea
where $e$ is the electric charge, while $\funciona$ and $\funcionb$ are   smooth functions.

When $N=3$,  the (time-independent)   scalar  $\psi$ and
  vector   $\>A$  potentials read 
$$ 
 \psi(\>q)=\funcionb(\>q^2)   -\frac{e}{2 }\,\>q^2
\funciona^2(\>q^2) 
 +\sum_{i=1}^3 \frac{\otra_i}{ 2   e q_i^2} \qquad \>A(\>q)= \>q \,\funciona(\>q^2) .
$$
Then the electric  $\>E=-\nabla \psi$ and magnetic $ \>H=\nabla\times \>A$  fields
turn out to be
$$ 
  \>E=\left(  e   \funciona^2+  {2e}  \>q^2 \funciona
\funciona^\prime-2\funcionb^\prime    \right)\>q  +\frac 1 {e} \biggl(\frac{\otra_1}{ 
q_1^3},\frac{\otra_2}{ 
q_2^3},\frac{\otra_3}{ 
q_3^3} \biggr) 
\qquad \>H=0 
\label{ch}
$$
where $\funciona^\prime$ and $\funcionb^\prime$ are
the derivatives with respect to the variable $\>q^2$.   
This kind of construction  
can also be  applied  to obtain certain
$N$D Fokker--Planck Hamiltonians  (see
\cite{Hietarinta} and references therein).


\subsection{Free motion on Riemannian spaces of constant curvature}

The  kinetic energy  ${\cal T}$ of a particle 
on the $N$D  Euclidean space ${{\mathbb E}^N}$
is just given by the generator
$J_+$ in the symplectic realization (\ref{be})   with {\em all} $\otra_i=0$:
\begin{equation}
 { H}  \equiv {\cal T}=\frac 12 \jp   =\frac 12\, {\>p}^2 .
 \label{bbff}
\end{equation}
Surprisingly enough, the kinetic energy 
  on $N$D Riemannian spaces with constant sectional curvature $\kappa$ can also be  expressed
in Hamiltonian form as a function of the $N$D symplectic realization of
the
$\frak{sl}(2,\R)$ generators  (\ref{be}).  In fact, this can be done in two different
ways
\cite{BHletter}:
\begin{equation}
\begin{array}{l}
\displaystyle{ { H} ^{\rm P}\equiv {\cal T}^{\rm P}=\frac{1}{2}\left( 1+\k
J_-   \right)^2 J_+  =
\frac{1}{2}\left( 1+\k \>q^2\right)^2 \>p^2} \\[8pt]
\displaystyle{{ H} ^{\rm B}\equiv {\cal T}^{\rm B} =\frac{1}{2}\left( 1+\k
J_-  \right)\left(  J_+   +\k J_3^2\right)=
\frac{1}{2}(1+\k \>q^2)\left( \>p^2+\k (\>q\cdot \>p)^2 \right) }.
\end{array}
\label{dd}
\end{equation}
In these expressions the function ${\cal T}^{\rm P}$ is just the kinetic energy for a
free particle  on  the spherical ${\mathbb S}^N$  
($\k>0$) and hyperbolic
${\mathbb H}^N$    ($\k<0$) spaces  when Poincar\'e
coordinates $\>q$ and their canonical momenta $\>p$ (coming from a stereographic projection in
${\mathbb R^{N+1}}$~\cite{Doub}) are used.  On the other hand
${\cal T}^{\rm B}$ corresponds to Beltrami coordinates and momenta (central projection).
The canonical transformation between both sets of phase space variables can be found in~\cite{BHKC}.

We can immediately conclude that, by construction, both Hamiltonians are QMS ones since they are   $\frak{sl}(2,\R)$-coalgebra symmetric. In other words, we can say that the $N$D Riemannian spaces with constant curvature are $\frak{sl}(2,\R)$-coalgebra spaces.


\subsection{Superintegrable potentials on spaces with constant\\
 curvature}

Without breaking the   $\frak{sl}(2,\R)$-coalgebra symmetry,
QMS potentials   on constant curvature spaces can now be 
constructed by adding some suitable functions depending on
$J_- $ to (\ref{dd}) and also by considering arbitrary centrifugal  terms that
come from symplectic realizations of the $J_+ $ generator with
generic $\otra_i$'s. With all these ingredients, the full $\frak{sl}(2,\R)$ Hamiltonian will be of the form
\begin{equation}
H ={\cal T}\left(J_- ,J_+ ,J_3 \right) + {\cal U}\left(J_- \right).
\end{equation}
The QMS systems  obtained in this way are just the curved counterpart of
the Euclidean systems, and by making use of the curvature parameter $\k$ we
can simultaneously describe the spherical
${\mathbb S}^N$  
($\k>0$),
hyperbolic ${\mathbb H}^N$    ($\k<0$) and Euclidean ${\mathbb E}^N$   
 ($\k=0$) cases. 

We remark that   special  choices for ${\cal U} $ lead to the many interesting QMS
systems on constant curvature spaces, that can  always be  expressed in both Poincar\'e   and Beltrami phase space coordinates.  In what follows we write three types of examples of this kind~\cite{BHletter} which 
share the {\em same} set of constants of the motion (\ref{cinfm}) 
although the geometric meaning of the canonical
coordinates and momenta can be different for each example.


\subsubsection{Curved Evans systems}
The $N$D constant curvature generalization
of the generic 3D Euclidean system with radial symmetry~\cite{Evansa} is
given by
\bea
&& { H} ^{\rm P}    ={\cal T}^{\rm P}+{\cal U} \left(   \frac{4
J_- }{ \left(1-\k J_- \right)^2 } 
\right)\nonumber\\ &&\quad\ \ =
\frac12 {\left( 1+\k \>q^2\right)^2 \>p^2}+{\cal U}\left(
\frac{4\>q^2}{(1-\k\>q^2)^2} \right) + \left( 1+\k \>q^2\right)^2 \sum_{i=1}^N\frac{\otra_i}{2 q_i^2}\label{eb}\\
&&
\displaystyle{ {  H}^{\rm B}={\cal T}^{\rm B}+ {\cal U}\left(J_-\right)=
\frac{1}{2}(1+\k \>q^2)\left( \>p^2+\k (\>q\cdot \>p)^2 \right) +{\cal U}\left(
\>q^2\right)+(1+\k \>q^2) \sum_{i=1}^N\frac{\otra_i}{2 q_i^2}}\nonumber
\eea
where ${\cal U}$ is  the smooth function that gives   the  central
potential and  its specific  dependence  in terms of
$J_-$   corresponds to the square of the radial distance in each coordinate
system~\cite{BHletter}. Hence, the Hamiltonians (\ref{eb}) are the curved generalization of the (flat) Evans systems on $\mathbb E^N$ (\ref{anha}), which are recovered under the limit (contraction) $\k\to 0$.


\subsubsection{The curved Smorodinsky--Winternitz  system}
This well-known Hamiltonian~\cite{KalninsH2,RS,PogosClass1,PogosClass2,VulpiLett,CRMVulpi} is just the Higgs
oscillator~\cite{Higgs,Leemon}  (that comes from the function ${\cal U}$
in (\ref{eb})) plus the corresponding $N$ centrifugal terms:
\begin{equation}
\begin{array}{l}
\displaystyle{ {  H}^{\rm P}={\cal T}^{\rm P}+   \frac{ \omega^2 
J_-}{ 2 (1-\k J_-)^2 } =
\frac12 {\left( 1+\k \>q^2\right)^2 \>p^2}}+ 
\frac{\omega^2 \>q^2}{2(1-\k\>q^2)^2}  + \left( 1+\k \>q^2\right)^2 \sum_{i=1}^N\frac{\otra_i}{2 q_i^2}
\\[8pt]
\displaystyle{ {  H}^{\rm B}={\cal T}^{\rm B}+\frac12 \omega^2 J_- =
\frac{1}{2}(1+\k \>q^2)\left( \>p^2+\k (\>q\cdot \>p)^2 \right) +\frac 12 \omega^2 
\>q^2 +(1+\k \>q^2) \sum_{i=1}^N\frac{\otra_i}{2 q_i^2}}.
\end{array}
\label{ec}
\end{equation}
This is a MS Hamiltonian whose integrals of the motion are quadratic in the momenta. The only
integral of the motion that does not come from the coalgebra symmetry has been explicitly written, in this variables, in~\cite{BHletter}. Clearly the limit $\k\to 0$ of both expressions (\ref{ec}) gives rise to the superposition of the $N$D isotropic harmonic oscillator with $N$ centrifugal terms in Cartesian coordinates:
$$
 H =\frac 12 J_+ +\frac12 \omega^2 J_- =
\frac{1}{2} \>p^2 +\frac 12 \omega^2 
\>q^2 +  \sum_{i=1}^N\frac{\otra_i}{2 q_i^2} .
$$


\subsubsection{The curved
generalized Kepler--Coulomb system}

The  Kepler--Coulomb potential on constant curvature spaces~\cite{RS,
PogosClass1,PogosClass2,Schrodinger,Schrodingerdual,
Schrodingerdualb,Miguel,Williams,kiev} with coupling constant $k$ and $N$ centrifugal terms 
is given by
\bea
&&
   {  H}^{\rm P}={\cal T}^{\rm P}-k \left(\frac{  
J_-}{ (1-\k J_-)^2 } \right)^{-1/2}\nonumber\\
&&\qquad \! =
\frac12 {\left( 1+\k \>q^2\right)^2 \>p^2 }- k \,
\frac{(1-\k\>q^2)}   {  |\>q| }+ \left( 1+\k \>q^2\right)^2 \sum_{i=1}^N\frac{\otra_i}{2 q_i^2} \label{ee}\\
&&
\displaystyle{ {  H}^{\rm B}={\cal T}^{\rm B} -k  J_-^{-1/2}=
\frac{1}{2}(1+\k \>q^2)\left( \>p^2+\k (\>q\cdot \>p)^2 \right) -\frac{k}{
 |\>q| } +(1+\k \>q^2) \sum_{i=1}^N\frac{\otra_i}{2 q_i^2}}.
\nonumber
\eea 
This has   recently been shown  to be a MS system~\cite{BHKC,Verrier}. When, at least, one  of the centrifugal terms vanishes ($\otra_i=0$) the  remaining  constant of the motion is quadratic in the momenta~\cite{BHletter}, whilst when {\em all} the constants $b_i\ne 0$, the additional integral is quartic in the momenta and has   been recently  presented in~\cite{BHKC}. We recall that the MS nature of the Euclidean case
$$
 {  H} = \frac 12 J_+ -k  J_-^{-1/2}=
\frac{1}{2}  \>p^2  -\frac{k}{
   |\>q| }  + \sum_{i=1}^N\frac{\otra_i}{2 q_i^2}
$$
was proven in~\cite{Verrier}, and it can be recovered from our formalism when $\k=0$.


\subsection{Free motion on spherically symmetric spaces of\\ nonconstant curvature}

With the previous examples in mind, it is easy to realize that any metric of the type
\begin{equation}\label{metric}
\d s^2=f(|\bq|)^2\d\bq^2\,
\end{equation}
leads to the free  Hamiltonian given by
\begin{equation}\label{H0}
H\equiv {\cal T}=\frac{\jp}{2 f(\sqrt{\jm})}=\frac{\bp^2}{2 f(|\bq|)^2}\,
\end{equation}
provided that all $b_i=0$. Thus this system determines the geodesic motion on the $N$D
spherically symmetric   space (\ref{metric}) with conformal factor $f(|\bq|)$;   this is generically a Riemannian space of nonconstant curvature. In particular, its scalar curvature  turns out to be
 \begin{equation}\label{curv}
R=-(N-1)\,\frac{2f''(|\bq|)+2(N-1)|\bq|^{-1}f'(|\bq|)+(N-2)f'(|\bq|)^2}{f(|\bq|)^2} 
\end{equation}
 where    $f'(|\bq|)=\dd f/\dd |\bq| $   and $f''(|\bq|)=\dd^2 f/\dd |\bq|^2$.

Therefore, since the Hamiltonian (\ref{H0}) 
  corresponds to an  $\frak{sl}(2,\R)$-coalgebra space 
then its geodesic flows define a QMS system for any choice of $f(|\bq|)$. 

Clearly, the geodesic flows on the Riemannian spaces of constant curvature described in section 4.3 are particular examples of the family (\ref{H0}) corresponding to set $f(|\bq|)=(1+\k \>q^2)^{-1}$ in Poincar\'e coordinates. Let us now describe other physically relevant examples with coalgebra symmetry.

\subsubsection{Darboux spaces}

Four interesting examples of this class of  spherically symmetric spaces with nonconstant curvature are the
$N$D generalizations of the four 2D Darboux spaces introduced in~\cite{BEHRdarboux}.  We recall that Darboux surfaces are the only 2D spaces
with nonconstant curvature for which there exists two functionally independent quadratic Killing tensors,
i.e., the geodesic motion on these 2D spaces is (quadratically) MS. There are only four spaces of
this type, that were characterized by Koenigs in a note included in the famous Darboux
treatise~\cite{Darboux}. It turns out that, by using appropriate charts, the four 2D Darboux metrics~\cite{darboux1,darboux2, pogoa, pogob} can
be rewritten in the form (\ref{metric}). Therefore, $N$D
spherically symmetric generalizations of such spaces can be obtained by expressing their kinetic energy
Hamlitonians with $\frak{sl}(2,\R)$-coalgebra symmetry. The four $N$D  Darboux metrics~\cite{BEHRdarboux} are explicitly written in table~\ref{table3} (where $a$ and $k$ are real constants).

Once again, we have that the $\frak{sl}(2,\R)$-coalgebra symmetry ensures that the geodesic  flows on all these spaces are QMS
Hamiltonian systems. In fact, one more independent integral of the motion is expected to exist for all of
them, since all these spaces are presumably MS in any dimension. Such an additional integral has already  been
found for the Darboux metric of type III~\cite{BEHRnd}.

\begin{table}[t]
{\footnotesize
 \noindent
\caption{{Examples of spherically symmetric spaces and their corresponding QMS free Hamiltonian with $\frak{sl}(2,\R)$-coalgebra symmetry.}}
\label{table3}
\medskip
\noindent\hfill
$$
\begin{array}{lll}
\hline
\\[-6pt]
\mbox{Space}&  \mbox{Metric} \  \dd s^2 &   \mbox{Hamiltonian}\ {\cal T}={\cal T}(J_-,J_+)\\[4pt]
 \hline
  \\[-6pt]
\mbox{Darboux I}&  \displaystyle {  \frac{\ln|\bq|\,\dd\bq^2}{\bq^2}}  & \displaystyle{\frac{J_- J_+}{2\ln\sqrt{J_-}}  }\\[10pt]
\mbox{Darboux II}&  \displaystyle {   \frac{1+ \ln^{2}|\bq| }{\bq^2\ln^{2}|\bq|}\, \dd\bq^2 }  & \displaystyle{\frac{J_- \ln^2\sqrt{J_-} \,J_+}{2(1+\ln^2\sqrt{J_-})}  }\\[10pt]
\mbox{Darboux IIIa}&  \displaystyle {      \frac{1+|\bq|}{\bq^4}\,\dd\bq^2 }  & \displaystyle{\frac{J_-^2 J_+}{2(1+\sqrt{J_-})} 
  }\\[8pt]
 {
\mbox{Darboux IIIb}}& { \displaystyle {  (k+ \bq^2)\,\dd\bq^2 }}  & {\displaystyle{ \frac{J_+}{2(k+ J_-)} 
  }}
  \\[10pt]
\mbox{Darboux IV}$\qquad$ &  \displaystyle {\frac{a+\cos(\ln|\bq|)}{\bq^2\sin^2(\ln|\bq|)}\,\dd\bq^2 \qquad}  & \displaystyle{\frac{J_-\sin^2(\ln\sqrt{J_-})J_+}{2(a+\cos(\ln\sqrt{J_-}))}
  }\\[12pt]
\hline
  \\[-8pt]

\mbox{Multifold Kepler}&  \displaystyle {  \frac{ (a+b |\bq|^{\frac 1\nu})\,\dd\bq^2}{   |\>q|^{2-\frac 1{\nu}}   }}  & \displaystyle{
\frac{ J_-^{1-\frac{1}{2\nu}} J_+}{2(a+b J_-^{\frac 1{2\nu}})}    }\\[12pt]
 \mbox{Taub-NUT}&   \displaystyle {  \frac{(4m+|\bq|)\,\dd\bq^2}{|\bq|}}  & \displaystyle{\frac{\sqrt{J_-} J_+}{2(4m+\sqrt{J_-})}    }\\[10pt]

 \hline
 \end{array}
$$
\hfill}
\end{table}
 

\subsubsection{ Iwai--Katayama spaces} These are the $N$D counterpart of the 3D spaces  
 underlying the so called  `multifold Kepler' systems introduced  in~\cite{IK94,IK95}, which depend on two real constants, $a$ and $b$, as well on a {\em rational} parameter $\nu$ as shown in table~\ref{table3}.   The physical interest of these systems relies on the fact that they are     generalizations of the   Taub-NUT  metric which   is recovered as the particular case with $\nu=1$,  $a=4m$ and $b=1$.  We stress that in the 3D case the multifold Kepler Hamiltonians  have been shown to be MS~\cite{CMP}, but  the additional integral of motion is not, in general, quadratic in the momenta.

\subsubsection{Potentials} We stress that centrifugal terms and central potentials can directly be added to the free system (\ref{H0}) by considering again an $N$D symplectic realization with arbitrary $b_i$'s plus a function ${\cal U}(\sqrt{\jm})$ as
\begin{equation}\label{H1}
H =\frac{\jp}{2 f(\sqrt{\jm})}+ {\cal U}(\sqrt{\jm})=\frac{\bp^2}{2 f(|\bq|)^2} + {\cal U}( |\>q|)+\frac 1{{2 f(|\bq|)^2} }\sum_{i=1}^N\frac{b_i}{q_i^2} .
\end{equation}
Several examples, such as oscillator and Kepler--Coulomb potentials on spherically symmetric spaces, can be found in~\cite{annals}.


\section{QMS Hamiltonians with $\frak{sl}_z(2,\R)$-coalgebra\\ symmetry}

A generalization of the construction
presented in the previous section  can be
obtained by making use of the non-standard quantum deformation  of  $\frak{sl}(2,\R)$ described in section 3.2 as
the dynamical symmetry   for the Hamiltonians corresponding to geodesic motion.  In this case,
the spaces so obtained are, in general, of nonconstant curvature, and the latter depends on the
deformation parameter $z$.

\subsection{{Free motion}}

In general, we can consider an infinite family of QMS geodesic flows with $\frak{sl}_z(2,\R)$-coalgebra symmetry (sharing the set of integrals (\ref{zcc}))
through the family of Hamiltonians 
\begin{equation}
{H}_z=\frac 12 \jp \, \g (z\jm )=\frac 12\, \g(z\>q^2)
\sum_{i=1}^N
 \frac {\sinh z q_i^2}{z q_i^2} \, p_i^2 \,    \exp\left\{ -
z\sum_{k=1}^{i-1}  q^2_k+ z
\sum_{l=i+1}^N q^2_l   \right\}  \label{bbfff}
\end{equation}
where $\g$ is a  smooth function such
that 
 $ {\lim_{z\to 0}\g\bigl(z\jm\bigr)=1}$, so that  $ \lim_{z\to 0}{H}_z =\frac 12 \>p^2$. This, in turn, means that $H_z$ is a QMS deformation of the free Euclidean motion which   defines a geodesic flow on an $N$D $\frak{sl}_z(2,\R)$-coalgebra
   space with   metric given by:
\begin{equation}
 \d s^2= \frac{1}{\g(z\>q^2)}
\sum_{i=1}^N
 \frac{2 z q_i^2} {\sinh z q_i^2} \, \d q_i^2 \,    \exp\left\{ 
z\sum_{k=1}^{i-1}  q^2_k- z
\sum_{l=i+1}^N q^2_l   \right\}   .
 \label{cc}
\end{equation}

For $N=2$, the  Gaussian curvature  of the space
 turns out to be~\cite{plb}  
 \be
 K(x)= {z}\left(  \g^\prime(x)\cosh x  +\left(
 \g^{\prime\prime}(x)-\g(x)-{\g^\prime}^2(x)/{\g(x)}
 \right) \sinh x
 \right) 
 \label{coo}
 \ee
 where $x\equiv z\jm=z(q_1^2+q_2^2)\equiv z\>q^2$, $\g^\prime=\frac{{\rm
 d}\g(x)}{{\rm d}
 x}$ and $\g^{\prime\prime}=\frac{{\rm d}^2f(x)}{{\rm d} x^2}$. For $N=3$, the   scalar curvature reads~\cite{yerevan}
\be
 R(x)=z\left( 6 \g^\prime(x)\cosh x  +\left(
  4   \g^{\prime\prime}(x)-5 \g(x)-5{ \g^\prime}^2(x)/ \g(x)
 \right) \sinh x
 \right) 
\label{co}
\ee
 where $x\equiv z\jm=z(q_1^2+q_2^2+q_3^2)\equiv z\>q^2$.

Now we briefly comment on some specific  $\frak{sl}_z(2,\R)$-coalgebra
   spaces that arise for simple choices of the function $g$ (see~\cite{plb,BHSIGMA,yerevan}). For each of them we write the Gaussian (\ref{coo}) and scalar (\ref{co}) curvatures corresponding to $N=2,3$.

 \begin{itemize}
 
 \item The simplest choice is to set $g(z J_-)=1$. The curvatures reduce to
 $$
 K = -   z \sinh\left(z \bq^2 \right) 
 \qquad  R = -  5 z \sinh\left(z\bq^2\right) 
 $$
 that is, they are   {nonconstant and {\em negative}} so the space is of hyperbolic type.

 \item Another possibility, which generalizes the above one, is   to take  $g(z J_-)=\exp(a z J_-)$ where $a$ is a real constant. The curvatures are given by
\bea
&&
 K =   z {\rm e}^{a z \>q^2} \left(  a \cosh\left(z \bq^2 \right) - \sinh\left(z \bq^2 \right) \right) 
 \nonumber\\
 && R=   z {\rm e}^{a z \>q^2} \left( 6 a \cosh\left(z \bq^2 \right) -(5+a^2) \sinh\left(z \bq^2 \right) \right) .
 \nonumber
\eea
It is worthy to remark that the very special cases with $a=\pm 1$ gives
$K=\pm z$ and $R=\pm 6 z$, so this  choice for $g$ includes the three classical Riemannian spaces of constant curvature (see section 4.3), for which the role of the sectional curvature $\kappa$ is now played by the real deformation parameter $z$.
 
  \item As a third example we consider $g(z J_-)=\cosh(z J_-)^b$ where $b$ is a real constant. This yields
 \bea
&&\!\!\!\!
 K =   z\left(  \cosh(z \>q^2)\right)^{b-2}  \sinh(z \>q^2) \left(  b +(b-1)  \cosh^2(z \>q^2) \right)  
 \nonumber\\
 &&\!\!\!\! R= -  z\left(  \cosh(z \>q^2)\right)^{b-2}  \sinh(z \>q^2) \left(  (5-10b)  \cosh^2(z \>q^2)  + b (4 +b)   \sinh^2(z \>q^2)  \right)  .
 \nonumber
\eea
When $b=1$,  the expressions are rather simplified as $K=z \tanh(z \>q^2)$ and $R=5 K$. 
 
 \end{itemize}
 
We stress that the geodesic motion of a classical particle on any of these spaces will have as integrals of the motion the functions (\ref{zcc}).


\subsection{Potentials}

We
can also introduce
more general $N$D QMS Hamiltonians based on $\frak{sl}_z(2,\R)$ (\ref{zsymp}) by
considering symplectic realizations with arbitrary
$b_i$'s (thus giving rise to `deformed centrifugal terms') and by adding   functions  depending on
$J_-$.  In particular, we have already studied the Hamiltonians  for the explicit 2D and 3D construction (see
\cite{jpa2D} and \cite{yerevan}, respectively). In this case the $N$D  Hamiltonian reads
\bea
&&
\!\!\!\!\!\! \!\!\!{ H}_z=\frac 12 \jp \, \g (z\jm 
 )+\pot (z\jm )  \label{ahaa}\\
 &&  =\frac 12\, \g(z\>q^2)
\sum_{i=1}^N \left( \frac {\sinh z q_i^2}{z q_i^2} \, p_i^2  +\frac{z \otra_i}{\sinh z
q_i^2} \right)    \exp\left\{ -
z\sum_{k=1}^{i-1}  q^2_k+ z
\sum_{l=i+1}^N q^2_l   \right\}  + \pot (z\>q^2) 
\nonumber
\eea
where the   smooth functions $\g$ and $\cal U$ are such that 
$$
\lim_{z\to 0}\pot(zJ_-)={\cal V}(J_-)\qquad 
\lim_{z\to
0}\g(zJ_-)=1.
$$
This condition means that the non-deformed/flat limit for these Hamiltonians is
$$
\lim_{z\to 0} { H}_z= \frac 12 \,\>p^2 + {\cal V}(\>q^2)
+\sum_{i=1}^N\frac{b_i}{2q_i^2},
$$
which are just the Evans systems (\ref{anha}).
So in this way we could define QMS analogues   of the
Smorodinsky--Winternitz and Kepler potentials~\cite{yerevan}  on the  $\frak{sl}_z(2,\R)$-coalgebra spaces.

In any case, a common feature of interacting systems with  $\frak{sl}_z(2,\R)$-coalgebra symmetry is the long-range nature of the
`interaction terms' depending on the coordinates and appearing in (\ref{ahaa}).


\section{Quasi-integrable Hamiltonians with $h_6$-coalgebra symmetry}

We present  three families of Hamiltonians with the underlying  $h_6$-coalgebra symmetry described in section 3.3. We recall that all of them are endowed with the  $2N-5$ integrals (\ref{c3m}) but, in principle,  they are only quasi-integrable since only $N-2$ integrals are in involution. Nevertheless, for some specific choices of the function $H$  the subalgebra structure of $h_6$ has allowed us to obtain the remaining integral by following the procedure already described in section 3.3. We omit here the explicit expressions for the families of integrable cases, that are fully described in~\cite{2photon}.

\subsection{Natural systems}

The Hamiltonian
\be
{H}=\frac 12 {B_+} +  {\cal F}\left( {A_-},B_-\right)
\label{eucl}
\ee
where ${\cal F}$ is a   function playing the role of a potential, gives rise to the following quasi-integrable system on   $\mathbb E^N$:
\be
{  H}=\frac 12 \>p^2+{\cal
F}\left(  \bla\cdot\>q, 
\>q^2\right).  
\label{eucla}
\ee
Notice that central potentials arise if $ {\cal F}$ does not depend on $A_-$;  in the case with generic $ {\cal F}\left( {A_-},B_-\right)$,    spherical symmetry is broken and QMS is, in principle, reduced to quasi-integrability.

\subsection{Electromagnetic Hamiltonians} 

The most general $h_6$-Hamiltonian including linear terms in the momenta is given by
\be
{ 
H}=\frac 12 {B_+}  + \kk\,{\cal F}\left( {A_-},B_-\right) + {A_+}\,{\cal G}\left(
{A_-},B_-\right) + \rrr\left( {A_-},B_-\right)
\ee
where   ${\cal F}$, ${\cal G}$ and $\rrr $ are   smooth functions. This means that
\be
  { 
H} =\frac 12 \>p^2 
+\left( \>q\cdot\>p -\frac {\bla^2}2\right) {\cal
F}\left(  \bla\cdot\>q,  \>q^2\right)   +\left( \bla\cdot\>p
\right) {\cal
G}\left(  \bla\cdot\>q,  \>q^2\right)  +
\rrr\left(  \bla\cdot\>q,  \>q^2\right) .
\label{electro}
\ee
In 3D, this Hamiltonian describes the motion of a particle on $\mathbb E^3$ under the action of a static electromagnetic field with  vector and scalar potentials given by
\bea
&&
\!\!\!\!\!\!\!\!\!\!\!\!\!\!
A_{i}=-\frac{q_{i}}{e}\mathcal{F}\left(A_{-},B_{-}\right)-\frac{\lambda_{i}}{e}\mathcal{G}\left(A_{-},B_{-}\right)\qquad i=1,2, 3
\nonumber \\
&& \!\!\!\!\!\!\!\!\!\!\!\!\!\!
\psi=\frac{1}{e}\mathcal{R}\left(A_{-},B_{-}\right)-\frac{1}{2e}M\,\mathcal{F}(A_{-},B_{-})
\nonumber\\
&& 
-\frac{1}{2e}\left[
B_{-}\mathcal{F}\left(A_{-},B_{-}\right)^{2}+2A_{-} \mathcal{F}(A_{-},B_{-})\mathcal{G}(A_{-},B_{-})+M\mathcal{G}(A_{-},B_{-})^{2}
\right] 
\eea
where $e$ is the electric charge. The $h_6$ quasi-integrability implies that, for any choice of the functions ${\cal F}$, ${\cal G}$ and $\rrr $, the Hamiltonian (\ref{electro}) commutes with the function
$$
C^{(3)}=
\left(
\la_1 (p_2 q_3- p_3 q_2 ) +
\la_2 (p_3 q_1 - p_1 q_3 ) +
\la_3 (p_1 q_2 - p_2 q_1 )
\right)^2.\nonumber\\
$$
Note that these systems can be thought of as a generalization of the electromagnetic Hamiltonians presented in section 4.2 and coming from the  ${\frak {sl}}(2,\mathbb R)$-coalgebra  symmetry (provided that all $b_i=0$).

\subsection{Geodesic flow Hamiltonians} 

A third family of relevant systems is  given by $N$D   Hamiltonians of the type 
$$
{  H}=\sum_{i,j=1}^N{g^{ij}(q_1,\dots,q_N)\,p_i\,p_j}
$$
that are obtained by considering 
\begin{eqnarray}
&& {  
H}=B_{+}\mathcal{F}(A_{-},B_{-})+A_{+}^2\mathcal{G}(A_{-},B_{-}) \nonumber \\
&& \qquad\qquad +\left(\kk+\frac{M}{2}\right)^{2} \rrr(A_{-},B_{-}) +
 \left(\kk+\frac{M}{2}\right)A_{+} \sss(A_{-},B_{-}) 
 \label{free}
\end{eqnarray}
since for any choice of the functions ${\cal F}$, ${\cal G}$, $\rrr $ and $\sss$ we obtain a Hamiltonian which is a quadratic homogeneous function in the momenta; namely,
\begin{eqnarray}
&&
{  H}=  \>p^2  
\mathcal{F}\left(  \bla\cdot\>q,  \>q^2\right)  +\left( \bla\cdot \>p\right)^{2}\mathcal{G}\left(  \bla\cdot\>q,  \>q^2\right)   \nonumber \\
&& \qquad \qquad
+\left( \>q\cdot\>p \right)^{2} \rrr\left(  \bla\cdot\>q,  \>q^2\right)  +\left( \>q\cdot\>p \right)\left( \bla\cdot \>p \right) \sss\left(  \bla\cdot\>q,  \>q^2\right) .
\label{geod}
\end{eqnarray}
The specific form of the metric $g^{ij}$ is so determined by   ${\cal F}$, ${\cal G}$, $\rrr $ and $\sss$ which, in general, give rise to an $N$D space of {\em nonconstant} curvature.
In any case, the set of constants of   motion (\ref{c3m})  is universal and does not depend on the specific choice of the   functions in the Hamiltonian. 

 Moreover, additional potentials on these $h_6$-coalgebra spaces can be naturally considered   by adding  functions  such as, e.g., ${\cal U}(A_-,B_-)$ to the free Hamiltonian (\ref{free}). In this way the natural  Euclidean systems (\ref{eucl}) can be generalized to the curved spaces defined through (\ref{free}) without breaking the quasi-integrability of the free Hamiltonian.


\section{Integrable Hamiltonians with other  coalgebra\\ symmetries}

For the sake of completeness, we briefly present some integrable  Hamiltonians which are based on different coalgebras to the above considered. We recall that a systematic study of Lie--Poisson coalgebras with dimensions 3, 4 and 5 has been performed in~\cite{alfonso}, and some more examples can be found in~\cite{Marmo, BRjmp, BRcluster}.


\subsection{The Calogero--Gaudin   system}
The Calogero--Gaudin (CG) Hamiltonian~\cite{CG,Gau}
\be
H=\sum_{1\le i<j}^{N}{2\,p_i\,p_j\,(1- \cos(q_i-q_j))}
\label{he}
\ee
was proven in~\cite{BCR,BR} to have an underlying $\frak{so}(2,1)$-coalgebra symmetry and its
constants of the motion were identified with the coproducts of the
$\frak{so}(2,1)$  Casimir under a certain symplectic realization. The generalized CG system
\be
H=\bigg(  \sum_{i=1}^{N}{p_i} \bigg)^2+ 
\frac{1}{2}\sum_{i,j=1}^{N}{p_i\,p_j\bigl((\kappa_i + \kappa_j)\cos
(q_i - q_j)
  -   (\kappa_i - \kappa_j)\cos (q_i +  q_j)\bigr)}
\label{qq1}
\ee
where $\kappa_i$ are free parameters,  was also proven to be completely
integrable in \cite{BR}.

We stress that, in general, the modification of the chosen symplectic realization drastically
changes the `shape' of the Hamiltonian through an associated canonical transformation. For instance, by using the
Gelfan'd--Dyson symplectic map the very same CG system reads~\cite{BRjmp}
\be
H=\sum_{1\le i<j}^{N} \left\{ -\,p_i\,p_j\,(q_i-q_j)^2 -
b\,(p_i-p_j)\,(q_i-q_j)\right\} + \frac{b^2}{4}\,N^2 
\label{hk}
\ee
where $b$ is a constant. On the other hand, the $N=2$ rational
Calogero--Moser Hamiltonian~\cite{CM} has also been proven to have $\frak{sl}(2,\R)$-coalgebra
symmetry \cite{BRjmp}.


\subsection{An integrable deformation of the CG system from $\frak{so}_z(2,1)$}

As we have pointed out previously, given a Hamiltonian with certain coalgebra symmetry any quantum deformation of the
underlying coalgebra provides an integrable deformation of the initial
Hamiltonian. This was explicitly shown for the first time by constructing
the following deformed CG system~\cite{BCR}:
\be
H_z=\sum_{1\le i<j}^{N}  2\,\pi_i\,\pi_j\left(1- \cos(q_i-q_j)\right) 
\label{lp}
\ee
where the non-local deformations of the momenta are
\be \pi_k=2\,\frac{\sinh (\tfrac{z}{2}
p_k)}{z}\, \prod_{i=1}^{k-1} {\rm e}^{-\tfrac{z}{2}p_i}
\, \prod_{j=k+1}^{N}{\rm e}^{\tfrac{z}{2}p_j} .
\nonumber
\ee
The corresponding constants of the motion come from the (deformed)
coproduct of the (deformed) Casimir of  $\frak{so}_z(2,1)$ (the standard deformation
of $\frak{so}(2,1)$) and, as expected, in the limit
$z\to 0$ we recover the `classical' CG system given in~(\ref{he}). We
remark that the quantum mechanical version of this deformed CG system has
been explicitly solved~\cite{MRjmp,MRjpa} and a `twisted' version of the Gaudin
magnet has already been introduced~\cite{Bal} through a quantum deformation
of the $\frak{gl}(2,\R)$-coalgebra.


\subsection{Systems defined on the harmonic oscillator coalgebra}

The Hamiltonian~\cite{C}
\be
 H=(\lambda+\mu)\,\sum_{i=1}^{N}{p_i} 
+ 2\,\mu\,\sum_{i<j}^{N}{\sqrt{p_i\,p_j}\,\cosh (q_i- q_j)}
\label{qq2}
\ee
where $\lambda$ and $\mu$ are real constants,
can also be  proven to be coalgebra invariant under the harmonic
oscillator coalgebra
$(h_4,\Delta)$. The quantum version of this Hamiltonian has also been
solved~\cite{CVD} and the system turns out to be equivalent to a chain of
coupled oscillators~\cite{CVD,Kar}.


\subsection{Ruijsenaars--Schneider-like systems from a quantum\\ Poincar\'e coalgebra}

Another interesting example of coalgebra-invariant system is the following
analogue~\cite{BR} of the Ruijsenaars--Schneider model~\cite{RuS}:
\be
H_z=\sum_{i=1}^{N}{\cosh \theta_i 
\, \exp{  \left( -\frac{z}{2}\left(\sum_{j=1}^{i-1}{q_j}\right) 
+\frac{z}{2}\left(\sum_{k=i+1}^{N}{q_k}\right) \right) } }
\label{qq2RS}
\ee
where $(q_i,\theta_i)$ are canonically conjugate variables such that $\{ q_i,\theta_j\}=\delta_{ij}$. This
completely integrable Hamiltonian was obtained by using the Poisson
analogue of the quantum deformation of the (1+1)D Poincar\'e
algebra introduced in~\cite{VK}.


\section{Generalizations of coalgebra symmetry}

So far we have shown how the coalgebra formalism provides an algebraic approximation to the integrability properties of Hamiltonian systems defined on a chain of $N$ copies, $A\otimes
A\otimes\dots\otimes A$,
of a given coalgebra $A$. Amongst the possible generalizations of this approach, two of them have already  been analysed. The first one deals with integrable systems defined on the $N$-chain $
V\otimes A\otimes A \otimes \dots \otimes A$ in which $V$ is an  {$A$-comodule
algebra}~\cite{Marmo, comodule}. The second extension of the formalism deals with the so-called {\em loop coproducts}~\cite{loop}, that generalize in terms of loop algebras all the algebraic machinery presented in this paper. As a concluding section of this contribution, we briefly sketch both approaches.


\subsection{Comodule algebra symmetry}

We recall that a (right)  {\em  coaction} of a coalgebra $(A,\Delta)$ 
on a vector  space $V$  is a linear map $\phi: V
\rightarrow V\otimes A$ such that  the following diagram is commutative: 
\begin{center}
\setlength{\unitlength}{1cm}
\begin{picture}(6,4)
\put(5,2){$V \otimes A \otimes A$}
\put(.1,2){$V$}
\put(2.0,.8){$V \otimes A$}
\put(2.0,3.3){$V \otimes A$}
\put(3.5,3.2){\vector(4,-3){1.1}}
\put(3.5,1){\vector(4,3){1.1}}
\put(0.7,1.9){\vector(4,-3){1.1}}
\put(0.7,2.3){\vector(4,3){1.1}}
\put(0.8,2.9){$\phi$}
\put(0.8,1.2){$\phi$}
\put(4.45,2.9){$\phi \otimes {\rm id}$}
\put(4.45,1.2){${\rm id}\otimes \Delta$}
\end{picture}
\end{center}

 If $V$ is an algebra, we shall say that $V$ is an  {$A$-comodule
algebra} if the coaction
$\phi$ is a homomorphism with respect to the product on the algebra $V$:
\begin{displaymath}
\phi(ab)=\phi(a)\,\phi(b) \qquad \forall a,b \in V.
\nonumber
\end{displaymath}

Moreover, if $V$ is a Poisson algebra and
\begin{displaymath}
\phi(\{ a,b\})=\{\phi(a),\phi(b)\}\qquad \forall a,b \in V
\nonumber
\end{displaymath}
we will say that $V$ is a {Poisson $A$-comodule algebra}.

It is straightforward to realize that this comodule structure can be used to mimic the coalgebra construction (see~\cite{Marmo,comodule}) in order to define {\em completely integrable systems} on
$$
V\otimes A\otimes A \otimes \dots \otimes A.
$$
However, in this case the  {superintegrability is lost} since  `right' integrals do not exist (the left-right symmetry of the chain of algebras has been broken).

Clearly, $A$ itself is an $A$-comodule algebra, with the coproduct $\Delta$ playing
the role of the coaction $\phi$, so that the comodule symmetry is indeed a generalization of the 
coalgebra one. 
In \cite{comodule}  some examples of classical and quantum integrable systems with comodule symmetry have been presented. 
We illustrate here one of them. As we said in section 3.3, the two photon algebra $h_6 =\langle \kk,A_+,A_-,B_+,B_-,M\rangle$ with Poisson brackets (\ref{aa}) contains the $\mathfrak {gl}(2,\R)=\langle B_-,B_+,\kk,M\rangle$
subalgebra. Moreover,  it is easy to realize that $\mathfrak {gl}(2,\R)$ is a (deformed) $h_6$-comodule algebra with the coaction $\phi$ defined by:
\begin{eqnarray}
&& \phi(M)=1 \otimes M+ M \otimes 1 \nonumber \\
&& \phi(\kk)=1 \otimes \kk + \kk \otimes \frac{1}{1-\sigma A_-} + M \otimes \frac{\sigma A_-}{1-\sigma A_-} \nonumber\\
&& \phi(B_+)=1 \otimes B_+ + B_+ \otimes \frac{1}{(1-\sigma A_-)^2}- 2 \sigma \kk \otimes \frac{A_+}{1-\sigma A_-}-
\sigma^2 \kk^2 \otimes \frac{M}{(1-\sigma A_-)^2} \nonumber\\
&& \phi(B_-)=1 \otimes A_-+A_- \otimes 1 - \sigma A_- \otimes A_- \label{cos}
\end{eqnarray} 
where $\sigma$ is a deformation parameter.
If we consider the following symplectic realization of $h_6$
\begin{eqnarray*}
D(B_+)=q_1^2 & D(B_-)=p_1^2 & D(K)=-p_1 q_1+\frac{\lambda_1^2}{2} \\
D(M)=-\lambda_1^2 & D(A_+)=-\lambda_1 q_1 & D(A_-)=-\lambda_1 p_1 
\end{eqnarray*}
and we take as the Hamiltonian on $\mathfrak {gl}(2,\R)$ the function
$$
H=\frac{1}{2} \left( B_+ + B_- \right)
$$  
the symplectic realization $D$ will give us the 1D
harmonic oscillator:
$$
H^{(1)}:=D(H)=\frac{p_1^2}{2}+\frac{q_1^2}{2}.
$$
Now, by using the coaction map (\ref{cos}) we can define a Hamiltonian
function on $\mathfrak {gl}(2,\R)\otimes h_6$:
$$
\phi(H)=\frac12 \left( \phi(B_+)+ \phi(B_-) \right).
$$
This Hamiltonian can be expressed in terms of canonical
coordinates by taking the symplectic realization
$D\otimes D$ of (\ref{cos}). It reads:
\begin{eqnarray*}
&& \back\back H^{(2)}_\sigma=(D\otimes D)(\phi(H))=\frac{1}{2}(p_1^2 + p_2^2)+\frac{q_2^2}{2}
+ \frac{q_1^2}{2(1+\sigma\,\lambda_2\,p_2)^2}
\cr
&& \qquad\quad + \sigma\,\lambda_2\left( p_1^2\,p_2 +
\frac{q_2(\lambda_1^2 - 2
q_1\,p_1)}{2(1+\sigma\,\lambda_2\,p_2)}\right)
+
\sigma^2\,\lambda_2^2\,\left(\frac12 p_1^2\,p_2^2 + 
\frac{(\lambda_1^2 - 2
q_1\,p_1)^2}{8(1+\sigma\,\lambda_2\,p_2)^2}   \right).
\label{h2s}
\end{eqnarray*}
This Hamiltonian is just an integrable deformation of the
2D
isotropic oscillator, since $\lim_{\sigma\to 0}{H^{(2)}_\sigma}=\tfrac{1}{2}(p_1^2 + p_2^2)+\tfrac12 (q_1^2 +
q_2^2)$. The integral of  motion in involution with $H^{(2)}_\sigma$ is just obtained as
the coaction of the $\mathfrak {gl}(2,\R)$ Casimir:
\bea
&& \back\back C^{(2)}_\sigma=(D\otimes D)(\phi(C))
\label{c2s}  \\
 && \qquad\, \back\back \! \!
=-\frac{\{ 2(p_2 q_1-p_1 q_2) +
\sigma p_1 (2 p_1 q_1- 4 p_2 q_2-\lambda_1^2) -
\sigma^2 \lambda_2^2 p_1 p_2 (-2 p_1 q_1+2  p_2 q_2+\lambda_1^2)
\}^2}{16\,(1+\sigma \lambda_2 p_2)^2} .
\nonumber
\eea
As expected, the limit $\sigma\rightarrow 0$ of (\ref{c2s}) is just
$-(p_2 q_1-p_1 q_2)^2/4$. Obviously, further iterations of the coaction map would 
provide the corresponding integrable deformation of the isotropic
oscillator in an arbitrary dimension. Nevertheless, the explicit form of such a Hamiltonian is quite involved due to the form of the coaction mapping $\phi$.


\subsection{Loop coproducts}

In \cite{loop} a further generalization of the coalgebra approach is proposed by weakening the Poisson homomorphism property:
$$
\Delta^{(m)}(\{a,b\})=\left\{ \Delta^{(m)}(a), \Delta^{(m)}(b) \right\}.
$$
Here we recall how the generalization works for the standard coalgebra structure (\ref{standardcop}) associated with a Lie--Poisson algebra.
  
Let $A$ be a Poisson algebra with generators $\{X^\al\},\ \al=1,\dots,l={\rm dim}(A)$, with Poisson brackets given by
$$
\left\{ X^\al, X^\beta \right\} = F^{\al \beta}( \vec{X})\qquad  \vec{X}=(X^1,\dots ,X^l)
$$
and with $r$ Casimirs functions ${\cal{C}}_j, \ j=1,\dots,r$. Let us assume that
one succeeds in finding a set of $m$ maps depending on a parameter $\la$:
\begin{equation}
\Delta^{(k)}_{\lambda}: A \to  A\otimes
A\otimes\dots^{N)}\otimes A    \qquad k=1, \dots, m \label{loopcop}
\end{equation}
such that for the generators of $A$ the following relations hold:
\begin{eqnarray}
&& \back \back \{ \Delta_{\la}^{(i)}(X^\al), \Delta_{\mu}^{(k)}(X^\beta) \}= f^{ik}(\la,\mu) F^{\al \beta} \left(\Delta_{\la}^{(i)}(\vec{X})\right)\qquad k>i \label{p1}\\
&& \back \back \{ \Delta_{\la}^{(k)}(X^\al), \Delta_{\mu}^{(k)}(X^\beta) \}=f^k(\la,\mu)F^{\al \beta} \left(\Delta_{\la}^{(k)}(\vec{X})\right)+   
g^k(\la,\mu) F^{\al \beta} \left(\Delta_{\mu}^{(k)}(\vec{X})\right) \label{p2}
\end{eqnarray}
where $f^{ik},f^k,g^k$ are some functions only depending on the parameters $\lambda$ and $\mu$. 
If the maps $\Delta^{(k)}_{\lambda}$ are defined on any smooth function $h$ of the generators of $A$ as: 
$$
\Delta^{(k)}_{\lambda} \left(h(X^1,\dots,X^l)\right) = h\left(\Delta^{(k)}_{\lambda}(X^1),\dots,\Delta^{(k)}_{\lambda}(X^l)\right),
$$
then it can be proven that the following relations are satisfied~\cite{loop}:
\begin{eqnarray}
&& \{ \Delta^{(i)}_{\lambda}({\cal{C}}_j) ,  \Delta^{(k)}_\mu (X^\beta) \} = 0 \qquad k > i  \label{lambda1} \\
&& \{ \Delta^{(i)}_{\lambda}({\cal{C}}_j) ,  \Delta^{(k)}_{\mu}({\cal{C}}_n) \} = 0  .\label{lambda2}
\end{eqnarray} 

This result can be interpreted as a generalization of the coalgebra approach. If $A$ is a Lie--Poisson algebra with   commutation rules
$$
\{  X^\al,  X^\beta \}= C^{\al \beta}_\ga \,  X^\ga
$$
and equipped with the standard coproduct (\ref{standardcop}), then
$$
\{ \Delta^{(i)}(X^\al), \Delta^{(k)}(X^\beta) \}= C^{\al \beta}_\ga \, \Delta^{(i)}(X^\ga)\qquad k \geq i,
$$
i.e., this is of the form (\ref{p1}) and (\ref{p2}) with 
$$
f^{ik}=f^k=1\qquad g^k=0 \qquad F^{\al \beta} \left(\Delta^{(i)}(\vec{X})\right)=C^{\al \beta}_\ga \, \Delta^{(i)}(X^\ga).
$$
Equations (\ref{lambda1}) and (\ref{lambda2}), 
in turn, generalize equations (\ref{za1}) and (\ref{cor3}). Consequently, the maps (\ref{loopcop}) can be properly called `loop coproducts'. 

In \cite{loop} it is shown that the loop coproducts given by
\begin{equation}
\Delta_\la^{(k)}(X^\al)=\frac{\Delta^{(k-1)}(X^\al)}{\la}+ \frac{X^\al_k}{\la-\epsilon} \qquad  k=2,\dots,N \label{loopcop2}
\end{equation}
with $\epsilon$ an arbitrary constant parameter, satisfy equations (\ref{p1}) and (\ref{p2}). From equation (\ref{lambda2}) it follows that the residues
of the functions $\Delta^{(k)}_\lambda({\cal{C}}_j)$ ($k=2,\dots,N, \ j=1,\dots,r$) are in involution. Furthermore this family of functions in involution contains the standard coproduct of the Casimirs $\Delta^{(k)}({\cal{C}}_j)$.

 Finally, it can be shown that if the Lie--Poisson algebra $A$ is a simple one and 
$$
\{I_1,\dots,I_s\} \qquad s=\frac{l-r}2 
$$
(to be compared with (\ref{condb})) define an integrable system on $A$, then the residues of the functions $$\Delta^{(k)}_\lambda({\cal{C}}_j)\qquad (k=2,\dots,N, \ j=1,\dots,r)$$ together with
the functions $$\Delta^{(N)}(I_1), \dots, \Delta^{(N)}(I_s)$$ define an integrable system on $ A\otimes
A\otimes\dots^{N)}\otimes A$. This last result solves for this particular case the integrability 
problem mentioned in section \ref{Symplectic} for the case of simple Lie--Poisson algebras.

 \section*{Acknowledgments}

This work was partially supported by the Spanish Ministerio de   Ciencia e Innovaci\'on    under grant     MTM2007-67389 (with EU-FEDER support), by the    Junta de Castilla y
Le\'on  (Project GR224)   and by INFN--CICyT.


\end{document}